\documentclass[letterpaper,11pt]{scrartcl}
\usepackage{amsmath,amsfonts,amsthm,color,amssymb,mathtools}
\usepackage{graphicx}
\usepackage{hyperref}
\usepackage{geometry}
\usepackage{lineno, blindtext}
\usepackage{natbib}

\theoremstyle{remark}

\usepackage[normalem]{ulem}
\usepackage{comment}
\usepackage{enumerate}
\usepackage{subfigure}
\usepackage{stfloats}
\usepackage{tikz}
\usepackage[framemethod=TikZ]{mdframed}
\usepackage{xcolor}
\usepackage{float}
\usepackage{bbm}
\usetikzlibrary{matrix,shapes,chains,positioning,decorations.pathreplacing,arrows}

\newcommand{\price}{P}
\newcommand{\payoff}{V}
\newcommand{\prob}{Pr}
\newcommand{\pdf}{p}
\newcommand{\thebeta}{\beta_s}

\usepackage{enumerate}

\newcommand{\be}{\begin{equation}}
\newcommand{\ee}{\end{equation}}

\usepackage[T1]{fontenc}
\usepackage{titling}

 \newcounter{example}
\renewcommand{\theexample}{\thesection.\arabic{example}}

\mdfdefinestyle{example}{%
    backgroundcolor=blue!8,
    linecolor=blue,
    outerlinewidth=2pt,
    leftline=false,rightline=false,
    skipabove=\baselineskip,
    skipbelow=\baselineskip,
    frametitle=\mbox{},
}
\newmdenv[%
    style=example,
    settings={\global\refstepcounter{example}},
    frametitlefont={\bfseries Example~\theexample\quad},
]{example}
\newmdenv[%
    style=example,
    frametitlefont={\bfseries Example~\quad},
]{example*}

\newmdenv[%
    backgroundcolor=red!8,
    linecolor=red,
    outerlinewidth=1pt,
    roundcorner=5mm,
    skipabove=\baselineskip,
    skipbelow=\baselineskip,
]{redbox}

\renewcommand{\vee}{\,\vert\,}

\title{A Unified Bayesian Framework for Pricing Catastrophe Bond Derivatives}

\author{Dixon Domfeh\thanks{Department of Computer Science, Georgia Institute of Technology, 266 Ferst Dr, Atlanta, GA 30313. Corresponding author. Email: dnkwantabisa3@gatech.edu.},
Arpita Chatterjee\thanks{Department of Mathematical Sciences, Georgia Southern University, Statesboro, GA 30458. Email: achatterjee@georgiasouthern.edu.}, and
 Matthew Dixon\thanks{Department of Applied Mathematics, Illinois Institute of Technology, 10 S Wabash Ave, Chicago, IL 60616. Email: matthew.dixon@iit.edu.}
}

\begin{document}
\maketitle

\begin{abstract}
%This paper deals with the main statistical steps involved in pricing catastrophe bond derivatives via a full Bayesian approach. 
% for uncertainty quantification which can be easily adapted under different data generation processes without loss of generality. We introduce a Bayesian framework to model the two fundamental sources of risks in CAT bond pricing -- catastrophe and interest rate risks.
%Moreover, we further account for parameter uncertainty through these models, which leads to more reliable CAT bond prices. 
Catastrophe (CAT) bond markets are incomplete and hence carry uncertainty in instrument pricing. As such various pricing approaches have been proposed, but none treat the uncertainty in catastrophe occurrences and interest rates in a sufficiently flexible and statistically reliable way within a unifying asset pricing framework. Consequently, little is known empirically about the expected risk-premia of CAT bonds. The primary contribution of this paper is to present a unified Bayesian CAT bond pricing framework based on uncertainty quantification of catastrophes and interest rates. Our framework allows for complex beliefs about catastrophe risks to capture the distinct and common patterns in catastrophe occurrences, and when combined with stochastic interest rates, yields a unified asset pricing approach with informative expected risk premia.  Specifically, using a modified collective risk model -- Dirichlet Prior-Hierarchical Bayesian Collective Risk Model (DP-HBCRM) framework -- we model catastrophe risk via a model-based clustering approach. Interest rate risk is modeled as a CIR process under the Bayesian approach. As a consequence of casting CAT pricing models into our framework, we evaluate the price and expected risk premia of various CAT bond contracts corresponding to clustering of catastrophe risk profiles. Numerical experiments show how these clusters reveal how CAT bond prices and expected risk premia relate to claim frequency and loss severity.
%Moreover, we further account for parameter uncertainty through these models, which leads to more reliable CAT bond prices. 
\end{abstract}



\section{Introduction}

Natural catastrophes are on the rise in recent decades, and (re-) insurance companies and countries have limited capacity to absorb a fraction of the catastrophe risk. Increasingly, insurers turn to  insurance-linked securities (ILSs) as an alternative risk transfer (ART) mechanism for raising additional capital. To meet this demand, catastrophe (CAT) bonds, an emerging type of ILS, serve as a contingent claim on a set of specified trigger mechanisms,  with a payoff linked to insurance risk. In other words, they are derivatives with the underlying claims process contingent on natural catastrophic events. Since their inception in 1994, CAT bonds and other ILSs have grown with remarkable success \citep{Lane2021}. With outstanding capital increasing steadily, the ILS market capitalization is estimated at  \$41.8 billion as of 2020, see\footnote{\href{https://www.artemis.bm/dashboard/}{https://www.artemis.bm/dashboard/}}. High demand and supply have ensured the exceptional high market. Insurers, countries, and regional governments on the sell side view CAT bonds issuance as a viable option to raise capital to expand their capacity to cope with the ever-increasing catastrophe risk due to climate change. CAT bonds also appeal to investors on the buy-side due to their zero beta (see, e.g., \cite{A17}, \cite{A7}). Usually, the underlying catastrophe risk in CAT bonds has a low correlation with the market, although this may not always be the case (see, e.g., \cite{A18}, \cite{A16}). Moreover, in a historically low-interest-rate environment, CAT bonds have exceptional appeal to investors due to their high yield. 

As the CAT market expands as an alternative investment for investors, it has become increasingly important for researchers and investors to understand how these financial instruments are priced by the (re-) insurer and the secondary market. The methods prescribed in the literature to price these instruments vary and are far from unified, even sometimes contradicting one another. One of the fundamental challenges in pricing is that CAT bonds are traded in an incomplete market and hence their prices do not admit a unique equivalent martingale measure. To be able to price CAT bonds like any other traditional derivative requires a complete market. An incomplete market presents many possible prices for an asset corresponding to different risk-neutral measures (\cite{A6}).

Various pricing frameworks have been proposed for CAT bonds in the primary and secondary markets in the literature. Some early works treat CAT bonds as zero-beta security. In other words, the correlation between the underlying catastrophe risk and the market is zero. Therefore, investors should earn a zero-risk premium (see, e.g., \cite{A17}, \cite{A8}, \cite{A6}, \cite{A31} and \cite{A37}. Proponents of this valuation approach follow \cite{A39}, who argues that localized jumps in an asset price may be due to asset-specific events and are uncorrelated with the market. This assumption allows for the valuation of CAT bond prices under risk-neutral measures. The important implication here is that the aggregate loss processes (i.e., intensity and severity of losses) retain their original distributional characteristics after been transformed from the physical probability measure to the risk-neutral measure (\cite{A31}). A similar argument supported by the zero-risk premium is that catastrophic risk has marginal impact on the overall economy and therefore does not pose a ``systematic risk'' to the market (\cite{A51}). However, there is empirical evidence to suggest that catastrophic events may in fact have a substantial systematic impact on the market (see, e.g., \cite{A18}). A case in point, the current COVID-19 outbreak (which can be considered a natural catastrophe) saw U.S. stocks tumble 11\% in five days, see\footnote{\href{http://www.bloomberg.com/news/articles/2020-02-27/stock-slump-to-extend-in-asia-on-virus-fears-markets-wrap}{http://www.bloomberg.com/news/articles/2020-02-27/stock-slump-to-extend-in-asia-on-virus-fears-markets-wrap}}. Pandemic Emergency Financing bonds issued worth \$425 million by the World Bank went into default when the World Health Organization declared the COVID-19 as a pandemic\footnote{\href{https://www.washingtonexaminer.com/news/425m-in-world-bank-catastrophe-bonds-set-to-default-if-coronavirus-declared-a-pandemic-by-june}{https://www.washingtonexaminer.com/news/425m-in-world-bank-catastrophe-bonds-set-to-default-if-coronavirus-declared-a-pandemic-by-June}}.

We can further survey the literature on CAT bond pricing under two separate categories : (i) Arbitrage portfolio theory (APT) based approaches, under the assumption of a complete market; \& (ii) Econometric approaches, which do not enforce arbitrage or require complete markets. Broadly, This paper partially bridges this literature,  primarily building on the former approach but incorporating modeling capabilities motivated by the many empirical findings reported by the latter. We briefly review each here in turn before focusing our attention on generalized pricing frameworks which are capable of adopting modeling assumptions motivated by the empirical literature.

There are several examples in the literature of pricing methodologies appropriate for complete markets. For example, \cite{A52} develops an arbitrage model (based on the \textit{Arbitrage Portfolio Theory (APT)}) to price ILSs which incorporate catastrophe events in a stochastic interest rate environment. Notwithstanding the incomplete market, \cite{A52} attempts to vindicate the arbitrage approach to pricing CAT bonds by asserting that the catastrophic jump risk index can be mimicked by instruments such as energy and power derivatives. While the replication of the interest rate dynamics is possible, the analogous assumption concerning catastrophe risk requires more justification. An extension of this pricing approach is afforded by \cite{A41}, who price payoffs as contingent on only catastrophe risk. \cite{A26} proposes a closed-form model consistent with APT on the assumption of an arbitrage-free LIBOR term structure of interest rates. 

An alternative but potentially more insightful approach than APT modeling, is an econometric approach. The econometric approach is usually applied to CAT bonds already trading in the secondary market. The earliest work can be attributed to \cite{A30}, who presents a two-factor model using 16 catastrophe bonds issued in 1999. The authors model the expected excess return (EER) as a function of the probability of first loss (PFL) and the conditional expected loss (CEL). The EER or risk premium is modeled with a Cobb-Douglas function to capture the asymmetrical nature of the catastrophe losses and PFL. The spread on the bond is expressed as the sum of EER and expected loss (EL) and the price of a CAT bond is LIBOR plus this spread. \cite{A30} is seminal as it develops the first model to characterize the behavior of the CAT bond market. Several works that follow proliferate this characterization by examining the impact of events or attributing CAT bond prices in primary and secondary markets to statistical or macroeconomic factors. For example, \cite{A1} examine the impact of hurricane Katrina on CAT bond prices using the theoretical framework of \cite{A30}. The parameters of Lane's pricing framework are estimated with treed Bayesian estimation. Their results show that during the 2005 hurricane season, the investment-grade rating increases the impact of the conditional expected loss. \cite{A18} examine how financial crises and natural catastrophes affect CAT prices based on bond-specific information and macroeconomic factors.

\cite{Lane2008} analyze data on 247 CAT bonds issued on the
capital markets between 1997 and 2008 to investigate how catastrophe risks are
priced. Their analysis reveals that catastrophe risk prices are a function of the underlying peril, the expected loss, the wider capital market cycle, and the risk profile of the transaction. \cite{A43} utilizes a generalized additive model to examine the factors that affect the CAT bond premiums on 192 CAT bonds launched between 2003 and 2008, and comes to similar conclusions, specifically pointing out other factors such as insurance underwriting cycles, rating class, issuer, catastrophe risk modeler, territory covered, and trigger type as relevant drivers of CAT premiums in the primary market. Most recently, \cite{A2} analyze data on CAT bonds between 1997 and December 2012 and tests several hypotheses on the factors that affect CAT premiums in the primary market and develops a robust forecasting model for predicting the CAT bond spreads.  \cite{A2} finds the expected loss is the largest driver of prices but comes to similar conclusions on which other factors are important. Finally, \cite{A21} develop a multi-factor spread model in a panel data setting to evaluate the drivers of CAT spreads in the secondary market. 

Another key direction in the ILS pricing literature is the use of probability transforms for arbitrage-free pricing. Probability transforms are distortion operators that combine both actuarial and financial pricing theory. One such distortion operator, widely used in ILS pricing, is the Wang Transform (see, e.g., \cite{A22}, \cite{A28}, \cite{A32}). However, \cite{A44} shows that the Wang transform cannot lead to consistent and arbitrage-free prices for a general stochastic process. As a result, he argues that it cannot be used as a universal framework for pricing insurance and financial risk. \cite{A53} introduces a universal framework (Wang transform) for pricing financial and insurance risk through a transfer and correlation measure that extends the CAPM to price all kinds of assets and liabilities. He broadens the CAPM to non-normally distributed risk to obtain a new parameter called the ``market price of risk'', which is analogous to the Sharpe ratio in the case of normally distributed returns. The two-factor Wang transform inflates probability densities to cope with adverse expectations while deflating them to accommodate favorable outcomes (Wang, 2000, 2004). In other words, it accounts for extreme tail risk in the probability distribution. This is akin to the ``volatility smile'' in option prices. As a result, it incorporates a form of risk adjustment.

%\MFD{Does this paper adopt and build-in on this two-step valuation approach? It's unclear why this paragraph exists.}
The methodological underpinnings of this paper can be traced to more recent developments towards a general valuation framework. For example, the two-step valuation approach of \cite{A45} introduces a general valuation framework that is consistent with both market prices and actuarial risk pricing principles. \cite{A10} propose a ``fair valuation'' approach (analogous to Pelsser and Stadje's system) which is both market-consistent and actuarial. Of particular relevance to this article, \cite{A51} develop a CAT bond pricing model based on the two-step valuation approach. Their approach bears some semblance to the APT and probability transform approach. They propose a product pricing measure that combines a distorted probability measure related to the underlying catastrophe risk and a risk-neutral probability measure incorporating interest rate risk. In their product measure model, the two sources of risks (i.e., catastrophe and interest rate risks) are modeled separately and integrated to form the pricing framework for the CAT bond. 

Beyond the literature on how to formulate a sufficiently generalized valuation frameworks suitable for CAT bond pricing and risk, lies questions of how to incorporate reliable models of catastrophe risk. In a frequentist setting, some researchers (e.g., \cite{A6}, \cite{A31}, \cite{A37}, \cite{A47}, \cite{A51}) have attempted to derive CAT bond pricing models with a link to industry loss trigger indices. The former approaches assume that all types of catastrophe risk data recorded on an industry loss index exhibits the same characteristics. While this assumption simplifies the catastrophe modeling approach, the results have limited practical implementation. CAT bond contracts are peril/catastrophe-specific. For example, a CAT bond contract may cover a single peril (like Tornadoes) or multiple perils (like Tornadoes, Severe storms, and Winter storms). Simply characterizing all types of catastrophes on an industry loss index, based on surpassing a minimum catastrophic loss cost, with the same risk profile is unsuitable for CAT bonds.

\subsection{Overview}
This contribution of this paper is to fill a void in the CAT bond pricing literature by not only tailoring risk profiles to specific perils but also  robustly accounting for uncertainty through a Bayesian framework which leads to reliable estimates of expected risk-premia of CAT bonds.

We build on the two-step valuation approach for CAT bond pricing \citep{A51} by treating perils listed on a loss index as having distinct or grouped characteristics, and hence different risk profiles\footnote{Usually loss indices only include catastrophic losses to the industry of at least \$25 million and affecting a significant number of insurers and policyholders.}. We propose an appropriate statistical model for each peril type (emphasizing their unique features) as well as their joint characteristics. Consequently, we can tailor the fair prices, present values of bond contracts, and their expected risk premia based on their group risk profile.

The Wang transform distorts the physical distribution to the risk-neutral distribution. Specifically, it transforms the physical probabilities of catastrophe risk to the risk-neutral probabilities. If there are multiple independent processes, however, the transform will result in different risk-neutral measures. A further contribution of this paper it to propose the entropy approach, transforming interest rate and physical catastrophe distributions into a single risk-neutral measure. 

%Therefore, the probability transform paragraph relates to the entropy approach we propose. But I think I failed to make that connection.

CAT bonds modeling involves contingent cash flows based on the probability that a rare event will occur within the life term of a bond contract. Such a probabilistic problem naturally yields itself to Bayesian modeling techniques, where  uncertainty is treated as a first-class citizen. However, at the time of writing this study, there are no published papers that apply Bayesian methods to CAT bond pricing to the best of our knowledge. Furthermore, while past research has investigated the effects of catastrophe risk on CAT bond prices in the secondary market (e.g., \cite{A30}, \cite{A43}, \cite{A1}, \cite{A18}, \cite{A2}), no empirical research has yet examined the expected risk-premium of CAT bonds.

\begin{figure}[!h]
\label{Figure1}
\centering
\hspace{-1cm}{\includegraphics*[width=7in, height=3.5in]{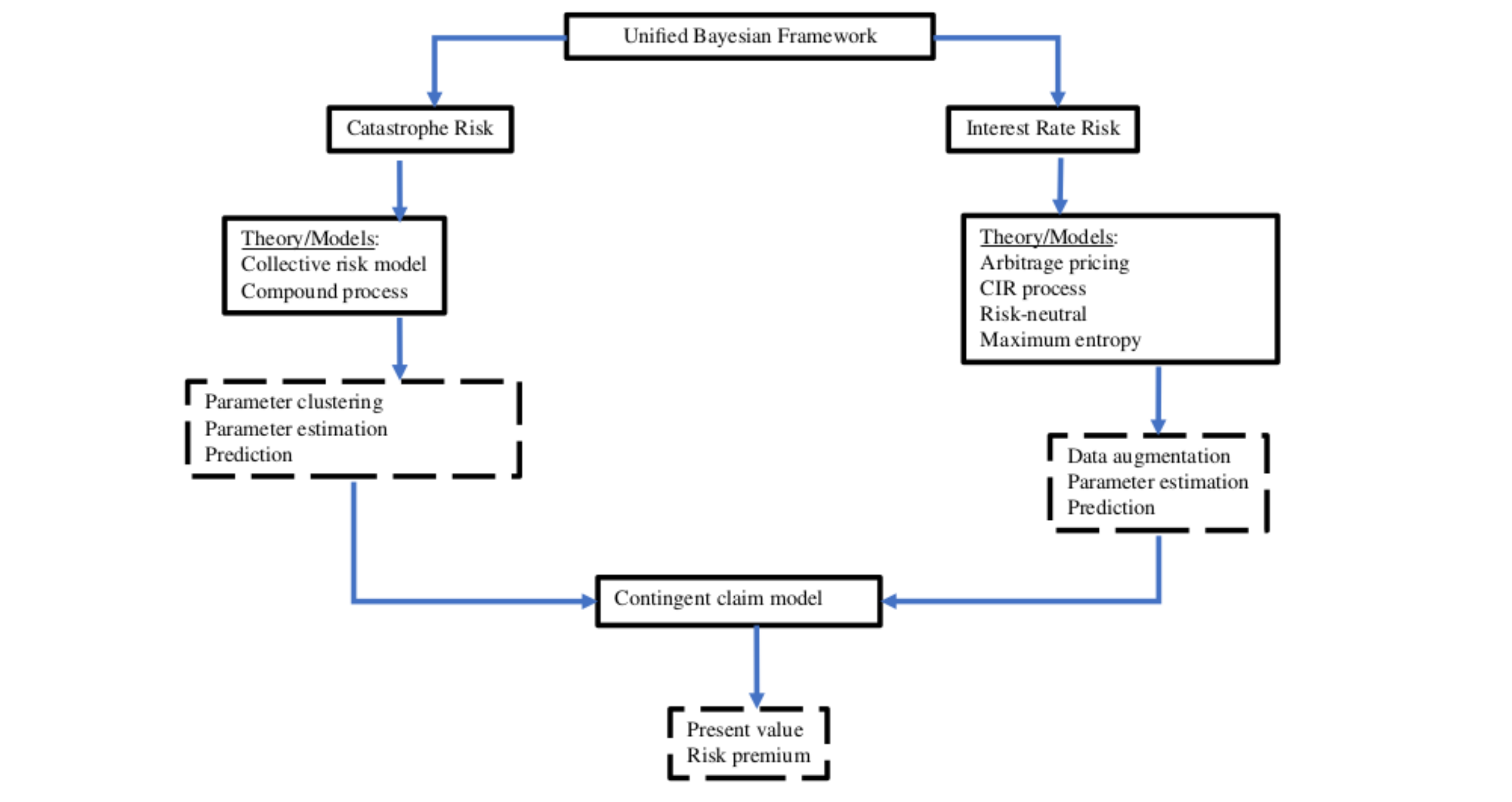}}
\vspace{-0.5cm}
\caption{A unified Bayesian framework for pricing CAT bonds}
\end{figure}

CAT bonds have two independent sources of risk -- catastrophe and interest risks. We develop two separate models for these risks and unify the results in our asset pricing framework (see Figure \ref{Figure1}). Following, \cite{A70}, we apply a hierarchical Bayesian collective risk model with a Dirichlet prior (DP-HBCRM) -- to model the catastrophe risk and a Cox Ingersoll Ross (CIR) model for the interest rate process. To assess the risk premia and present value of CAT contracts corresponding to different risk profiles, we utilize the entropy maximization principle to derive risk-neutral measures. We demonstrate the use of our model through several numerical examples. 
Our proposed framework reveals four distinct clusters of catastrophe risk profiles and we observe higher CAT bond values for clustered perils with a relatively lower number of claims. We also notice that risk groups with a higher probability of breaching a set loss threshold have higher expected risk premiums.\\

The remainder of this paper consists of six sections. We give a detailed outline of our Bayesian modeling approach to catastrophe risk in Section~\ref{sec2}. Then in Section~\ref{sec3}, we discuss our stochastic interest rate modeling methodology. The product probability measure model for contingent claim pricing is detailed in Section~\ref{sec4}. To assess the risk premia, we discuss and implement the entropy maximization principle and provide numerical examples in Section~\ref{sec5}. Finally, Section~\ref{sec6} concludes with suggestions for further research. 

%Such a portfolio has two primary sources of randomness -- the claim number process and the claim size process.
\section{Bayesian Collective Risk Model}\label{sec2}
\subsection{General Collective Risk Model}
%Such a portfolio has two primary sources of randomness -- the claim number process and the claim size process.
Our setup considers an industry trigger loss index as a portfolio of catastrophe risks that needs to be modeled.  Consistent with the classical actuarial literature, we characterize the catastrophe risks by two independent stochastic processes: (i) the number of claims (intensity) -- which counts the claims, and; (ii) the claim amounts (severity) -- which determines the monetary cost when a claim is reported.  The aggregate claims process is, therefore, a compound process of these two sources of risks.  We denote \textit{i}  as the peril types,  $i=1,2,\dots,I$ and denote $t=1,\dots,T$ the time interval for the arrival of a catastrophe event at quarterly intervals. We assume $N_{i,t}$ (with \textit{N}(0)= 0) is the random number of claims within an arbitrary time interval  $[0,\ t]$ with corresponding $k^{th}$ claim sizes as, $X_{i,k}\ k\mathrm{\ =\ 1,\ 2\ ,\dots \ }$ for each peril type. We further assume that the sequences $\left\{X_{i,1},X_{i,2},\ \dots \right\}$ are i.i.d. non-negative random variables with a common distribution $F\left(x\right)=\prob\left[X\le x\right]$. For simplicity, we treat the two processes $N_{i,t}$ and $X_{i,t}$ as independent for all $i$ and $t$. For each of these $N_{i,t}$ claims, an insurance company also records the claim amounts, $\left\{X_{i,1},X_{i,2},\ \dots \ \right\}$ generated. Therefore, the aggregate claims up to a fixed time $t\ge 0$ constitute the collective risk 
\begin{equation}\label{eq:2.1}
S_{i,t}=X_{i,1}+\dots +X_{i,N\left(t\right)}=\sum^{N_{i,t}}_{j=1}{X_{t,i,j}},\ \ t\ge 0,\ i=1,2,\dots ,I\ ,\ j=1,\dots,N, 
\end{equation}
with (\textit{S(t)} = 0 if \textit{N(t)} = 0). The sequence ${\left\{S_{i,t}\right\}}_{t\ge 0}$ is a stochastic process which is referred to as a compound or random sum. The moment generating function of $S$ conditioned on $N$ is $m_S\left(t\right)=\mathbb{E}\left[e^{tS}\right]=m_N\left(\log m_X(t)\right)$. For the distribution of $S$, one can find 
\begin{equation}\label{eq:2.2}
F_S\left(x\right)=\prob\left[S\le x\right]=\sum^{\infty }_{n=0}{\prob\left[S\le x \vee N=n\right]\cdot \prob\left[N=n\right]}=\sum^{\infty }_{n=0}{F^{n*}_X\prob\left[N=n\right]},
\end{equation}
where $F^{n*}_X= Pr[X_1+X_2+\dots +X_n\le x]$ is the n-fold convolution of the distribution of claim sizes.

\subsection{Bayesian Collective Risk Model}

The independence assumption in the general collective risk process may be too restrictive in many applications. For instance, some catastrophes such as seasonal tornadoes in North America follow a pattern, and as such, claim amount and claim numbers may not necessarily be independent. In such a situation, the i.i.d. assumption needs to be relaxed. For example, \cite{A47} show that by allowing for dependency between claim amounts and the inter-arrival times in the claims process via a semi-Markov risk model, one can obtain fairer CAT bond prices. \\

We argue that the catastrophe risk measure should include additional risk information about the process other than intensity and severity. CAT bond derivatives are often priced using historical data on insured catastrophe loss indices that contain different peril types. An example of such a data set is the catastrophe loss index of the Property Claim Service (PCS) which comprises \textit{all} catastrophe events in the U.S. with an insured cost of more than USD 25M. Such a loss index has heterogeneous aggregate losses due to the natural grouping of different peril types. Modeling would serve practical purposes if one could price CAT bonds based on their individual or grouped risk profiles. Furthermore, CAT bonds contract designs have become sophisticated. Insurers and re-insurers bundle multiple catastrophe risks (e.g., tropical cyclones, earthquakes, wildfire) in one bond contract. The assessment of the probability of a trigger event of such multi-peril bonds is an uneasy feat. These bond contracts may mask high layers of risk in CAT bond prices. It is increasingly difficult for investors to assess the fair value of risk of buying such financial instruments. \\

\begin{figure}[!h]
\centering
\includegraphics*[width=0.7\textwidth, height=0.45\textheight]{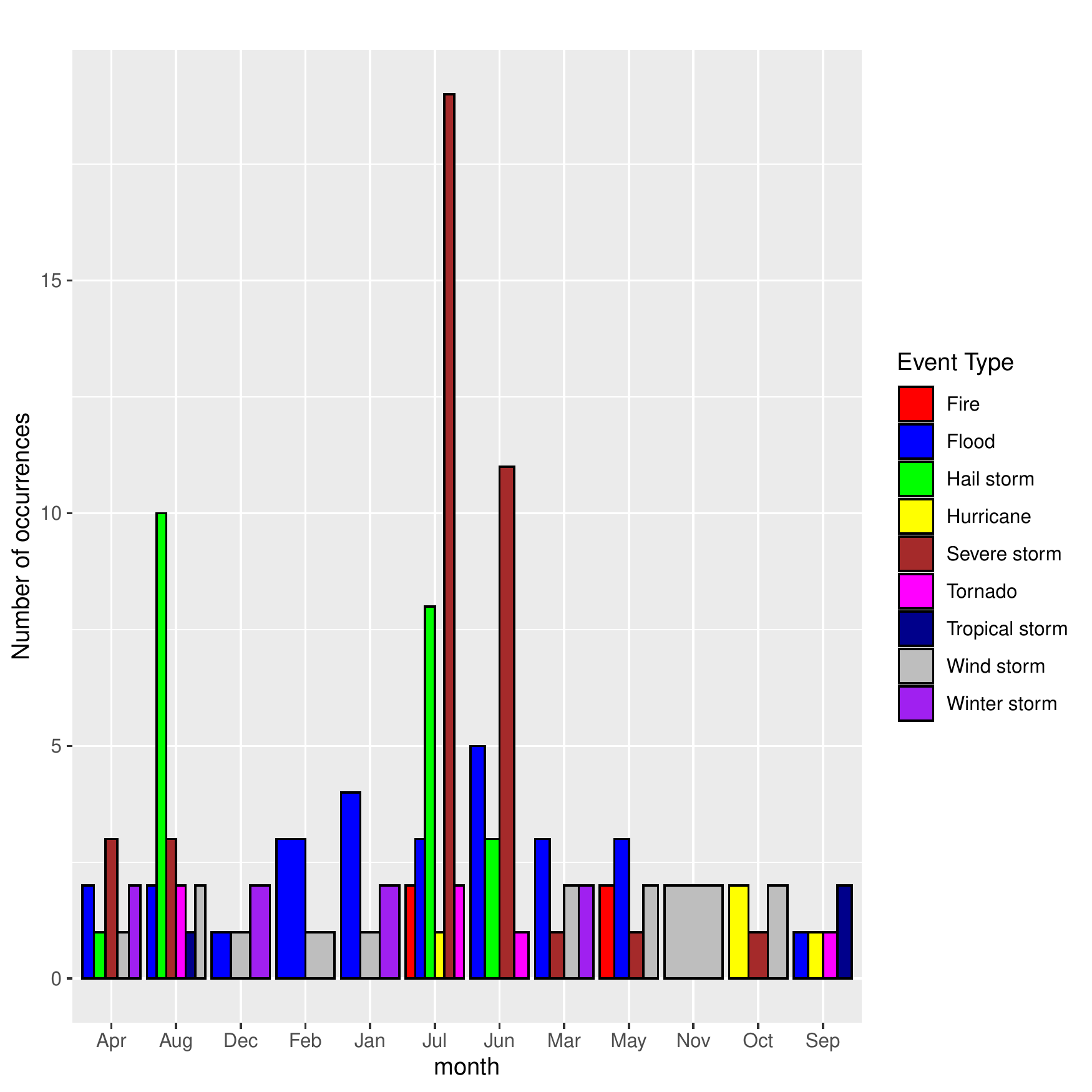}
% Figure 2.1: 
\caption{Seasonal patterns in catastrophic events by peril types.}
\label{Figure2.1}
\end{figure}

\noindent \textbf{Claim Number Process}: Several distributions can be used to model $N_{i,t}$ such as the Poisson, Negative Binomial, or Gamma. This paper adopts an extension of the Poisson process (i.e., nonhomogeneous Poisson process) to accommodate dependency of peril-specific characteristics in catastrophe frequency. We explicitly incorporate peril- and seasonal-dependency in the Poisson process via a likelihood specification of the form
\begin{equation*}
\begin{split}
& N_{i,s} \vee \lambda_{i,s} \sim Poisson(\lambda _{i,s}), \quad \lambda_{i,s}>0 \\
&\log(\lambda_{i,s}) \vee \alpha_i,\thebeta, X_s = \alpha _i+\thebeta X_s,  
\end{split}
\end{equation*}
where ${\lambda }_{i,s}$ is the average number of claims per quarter for particular perils for a fixed unit of time. The intensity of the arrival process, ${\lambda }_{i,s}$ is modeled as a Poisson regression. The motivation for this likelihood is that we want to capture the seasonality and peril-specific characteristics that may influence the final aggregate loss for different perils. Notably, the random intercept, ${\alpha }_i$ allows claims intensity to be correlated through peril type\footnote{ ${\alpha }_i$ is a peril-specific parameter which indicates the average intensity of each peril type   $i$, even if we ignore seasonality effects.}. The quarterly seasonal patterns of catastrophes are directly built into the model via, $X_s$. This approach is appropriate since the underlying risk in our portfolio of catastrophes shows discernible patterns. Some perils exhibit higher occurrence relative to others from June through August, as shown in Figure~\ref{Figure2.1}. 
\newpage 
\noindent\textbf{Claim Size Process}: To model the claim amount involves selecting an appropriate distribution for $F\left(x\right)$. One has to fit several probability density functions to the observed loss data and choose the distributions that best describe the data generating process. In actuarial science, an extreme or rare event occurs with a low probability but causes severe damage. Distributions with long- and heavy-tails are typically used to model these phenomena to assess tail risks. Since the choice of distribution can influence the CAT bond prices, we consider several heavy-tailed distributions (within the domain${\mathrm{\ }\mathbb{R}}_+$) via nonparametric tests. We utilize the Kolmogorov-Smirnov (K-S), and Anderson-Darling (A-D) tests for our model selection (see, e.g., \cite{A37}). For a random sample, the goodness of fit exercise is to test, $H_0$: the sample is drawn from a population with a theoretical distribution $G\left(x\right)$. \\

The K-S (\cite{A4}) test is technically distribution-free and nonparametric. In other words, K-S does not assume the distribution of the data. The K-S is used to test if a sample comes from a population of a specified distribution. Given \textit{N} ordered random samples $Y_1,Y_2,\dots,Y_N$, the empirical cumulative distribution function (ECDF) is defined as $F_N\left(x\right)=N^{-1}\times \left[number\ of\ observations\le x\right]$. The K-S statistic $D$, is based on the maximum distance between $F_N\left(x\right)$ and $G\left(x\right)$ for all values of $x$. The closer the $D$ statistic is to 0, the more likely the two samples are drawn from the same distribution. When comparing across several distributions, the distribution with the largest p-value is the best fit. The A-D (\cite{A49}) test modifies the K-S test and gives more weight to the distribution's tails (by using a squared distance) than the K-S test. Further, we assess the performance of the selected distributions, with the best fit model having the lowest AIC and BIC values. Tables \ref{Table2.1} and \ref{Table2.2} report these test statistics and the performance of the different distributions tested.

\begin{table}[!h]
%Table 2.1:  
\centering
\begin{tabular}{|p{0.8in}|p{0.8in}|p{0.9in}|p{0.8in}|p{0.8in}|p{0.8in}|} \hline 
Distribution & Weibull & Inv. Gamma & Pareto & Log-normal & Gamma \\ \hline 
\multicolumn{6}{|p{2.5in}|}{Test statistic (critical value = 0.05)} \\ \hline 
$D_n$ & 0.69271\newline $\left(2.2\times {10}^{-16}\right)$ & \textbf{0.088842\newline }$\left(0.2687\right)$\textbf{} & 0.21191\newline $\left(2.22\times {10}^{-6}\right)$ & 0.14371\newline $\left(0.01054\right)$ & 0.23535\newline $\left(1.55\times {10}^{-6}\right)$ \\ \hline 
A-D & 1.9341\newline $\left(0.000549\right)$ & \textbf{3.0864\newline }$\left(0.2472\right)$\textbf{} & 2.4579\newline $\left(0.4522\right)$ & 2.3471\newline $\left(0.4979\right)$ & 3.6679\newline $\left(0.1356\right)$ \\ \hline 
\end{tabular}
\caption{\textit{Test statistics for different distributions (p-values are in parenthesis).}}
\label{Table2.1}
\end{table}

%\noindent Table 2.2:
\begin{table}[!h]
\centering
\begin{tabular}{|p{0.8in}|p{0.8in}|p{0.9in}|p{0.8in}|p{0.8in}|p{0.8in}|} \hline 
\multicolumn{6}{|p{1in}|}{Distribution} \\ \hline 
Criteria & Weibull & Inv. Gamma & Pareto & Log-normal & Gamma \\ \hline 
AIC & 984.5139 & \textbf{889.1509} & 955.1093 & 917.1459 & 993.571 \\ \hline 
BIC & 990.2023 & \textbf{894.8393} & 960.7977 & 922.8343 & 999.2594 \\ \hline 
\end{tabular}
\caption{\textit{Performance of different distributions. Note: optimum values are in boldface. }}
\label{Table2.2}
\end{table}\newpage 

\noindent From Table \ref{Table2.1}, we fail to reject the $H_0$ that our data is drawn from an Inverse-Gamma distribution for the K-S test. The A-D test, with its emphasis on the tails of the distributions, suggests that all distributions can be a good fit based on all p-values $>$ 0.05. However, in terms of performance the Inverse-Gamma distribution shows the best fit with the lowest AIC and BIC values (see Table \ref{Table2.2}). Based on the model selection choice, the likelihood for claim size can be written as 
\begin{equation*}
X_{i} \vee \kappa_i, \theta_i \sim Inv.Gamma\left(\kappa_i,\theta_i\right),
\end{equation*}
where ${\ \theta }_i>0,\ \ {\kappa }_i>0$ indicate the shape and rate parameters.\\

\noindent \textbf{Aggregate Claim Process}: The random sum of individual claim sizes, $X_{i\ }$ conditioned on $N_{i\ }$as shown in Equation~\eqref{eq:2.1}, constitutes the total claims cost. For a fixed unit of time $t$, peril type $i$ and season-quarter $s$, knowing that $N_{i,s}=\ n_{i,s}$, the claim sizes $X_{i,s,j},\ j=1,2,\dots ,n_{i,s}$ are i.i.d. It follows that the sum of these inverse gamma distributions is also an inverse gamma distribution of the form
\begin{equation*}
S_{i,s} \vee n_{i,s}, \theta_i \sim Inv.Gamma\left( n_{i,s}\cdot \kappa_i,\theta _i\right),\quad \theta_i>0, \kappa_i>0. 
\end{equation*}

%2.2.1 
\subsubsection{Hierarchical Bayesian Collective Risk Model}
Hierarchical Bayesian models have several applications in insurance industry (see for example (\cite{A40}, \cite{A15}). A hierarchical Bayesian model shares information or ``borrows statistical strength'' from all levels of the data used. In other words, the hierarchical structure enables information flow among all peril types. This multilevel interaction and dependency throughout the hierarchy can ensure more reliable and robust estimates of model parameters. We create a hierarchical model by placing shared priors over parameters and estimate them directly from data. The likelihood of hierarchical collective risk model takes the form
\begin{equation}
\begin{split}
S_{i,s} \vee &n_{i,s}, \kappa_i, \theta_i \sim Inv.Gamma\left (n_{i,s}\cdot \kappa_i,\theta_i\right), \quad \theta_i>0, \kappa_i>0\\
 N_{i,s} &\vee \lambda_{i,s} \sim Poisson\left(\lambda_{i,s}\right),\quad \lambda_{i,s}>0\\
& \log \left(\lambda_{i,s}\right) \vee \alpha_i, \thebeta, X_s = \alpha_i+\thebeta X_s.  
\end{split}
\end{equation}
%2.2.2 
\subsubsection{Hierarchical Dirichlet Prior Specification}
We seek to impose the belief that some catastrophes may share similar characteristics in terms of their frequency of occurrence and claim amount, thus, forming clusters or groups. In this setting, it is natural to consider a Dirichlet Process (DP) prior, which induces partitions in the parameter space. As introduced by \cite{A14}, a Dirichlet process is a prior measure on the space of probability measures. The Dirichlet process $DP(\gamma ,\ G_0)$ has two parameters, a scale parameter, $\gamma >0,$ and a base probability measure${,G}_0$.\\

Consider $B$ as a measurable subset under, $G_0$ and $G\ \sim \ DP(\gamma ,\ G_0)$. The DP is formally defined by the property that, for any finite partition, $\left\{B_1,\ \dots ,B_k\right\}$ of the base measure $G_0$, the joint distribution of the random vector $(G\left(B_1\right),\ \dots ,\ G\left(B_k\right))$ is the Dirichlet distribution with parameters, $\left(\gamma G_0\left(B_1\right),\dots ,\gamma G_0\left(B_k\right)\right)$. The expectation and variance of $B$ is given as
\begin{equation*}
\mathbb{E}\left[G(B)\right]=G_0(B) \quad \text{and}\quad \mathbb{V}\left[G(B)\right]=\frac{G_0(B)\left\{1-G_0(B)\right\}}{\gamma +1}.
\end{equation*}
This implies that, $G_0$ is the mean, and $\gamma $ controls the concentration around the mean (\cite{A25}). An important property of DP is its large \textit{weak} support (\cite{A36}, pp. 8). The weak support suggests that \textit{any }distribution having the same support as $G_0,$ can be approximated \textit{weakly }by a DP probability measure. This property makes the DP prior one of the desirable processes in nonparametric Bayesian analysis. Another unique property of the DP is the discrete nature of $G$. With probability 1, we can always write $G\ \sim \ DP(\gamma ,\ G_0)$ as a weighted sum of point-masses,
\begin{equation*}
G=\sum^{\infty }_{h=1}{w_h{\delta }_{m_h}(\cdot )},
\end{equation*}
where $w_1$, $w_2,\dots $ are random weights and ${\delta }_x(\cdot )$ is the point-mass distribution at $x$ for random locations, $m_1,m_2,\dots \ $. The random weights, $w_1$, $w_2,\dots $ can be explicitly represented in \textit{the stick-breaking construction} outlined by \cite{A46} as
\begin{equation*}
w_1=v_1, \quad w_h=v_h\prod_{k<h}{\left(1-v_k\right)}, \quad \text{where}\quad  v_1,v_2,\dots \sim Beta\left(1,\gamma \right).
\end{equation*}
The above expression satisfies $\sum^{\infty }_{j=1}{w_h=1}$ with probability 1. Thus, the random weight, $w_h$ is a probability measure with positive integers. The stick breaking construction of a Dirichlet distribution is analogous to the running of polya-urn scheme by Blackwell and MacQueen (1973) indefinitely. We start with a stick of length one. We generate a random variable  $v_1 \sim Beta\left(1,\gamma \right)$ with expected value $1/(1+\gamma)$. We break off the stick at $v_1,$ and the length of the stick on the left is $w_1$. Now, we take the stick to the right, and generate $v_2 \sim Beta\left(1,\gamma \right)$, break off the stick at $v_2$ with $w_2=(1-v_1)v_2$ as the length of the stick to the left. We continue this process nonstop and this generates the stick breaking construction of the Dirichlet distribution. \\

In the implementation of the Dirichlet prior via the stick breaking construction, we set $\gamma =9$ which represents the number of peril types in our data set. Our initial assumption is that each peril type in terms of its number of claims and claim amount are unique. This \textit{a priori} assumption is very conservative, as it restricts our influence of the number of clusters that will be formed \textit{a posteriori} after running the hierarchical Bayesian model.\\

We specify two DP priors with different base distributions to induce separate partitioning on the parameter space by aggregate claims and claim numbers\footnote{ Two DP priors are needed due to the independent relationship between claim amount and claim number processes.}. Specifically, to ensure that model parameters are grouped for multiple perils which share common aggregate claim amounts, but different claim count characteristics, we  first introduce 
\begin{equation*}
\begin{split}
 &(\kappa_i, \theta_i)\ \sim \ DP({\gamma }_1,\ G_0(\cdot )),\\ 
& G_0\left({\zeta }_1,{\zeta }_2,{\eta }_1,{\eta }_2\right)=Gamma({\zeta }_1,{\zeta }_2)\times Gamma({\eta }_1,{\eta }_2), 
\end{split}
\end{equation*}
where ${\gamma }_1$, are the concentration parameters, $G_0\left(\cdot \right)$ is the base distribution of the Dirichlet prior and  ${\zeta }_1,{\zeta }_2,{\eta }_1,{\eta }_2$ are hyper-parameters. The shape and scale parameters ${(\kappa }_i,{\theta }_i)$ are modeled as a bi-variate gamma distribution with  $\left({\zeta }_1,{\zeta }_2,{\eta }_1,{\eta }_2\right)$ as hyper-parameters. $Gamma({\zeta }_1,{\zeta }_2)$ controls the shape while $Gamma({\psi }_1,{\psi }_2)$ controls the scale. The approach provides much flexibility in the model to account for the fact that different perils may exhibit other characteristics. Further, we induce clusters in the claims number process by introducing $Gamma\left({\psi }_1,{\psi }_2\right)$ over${\ \alpha }_i$ as
\begin{equation*}
\begin{split}
&\alpha_i  \sim \ DP({\gamma }_2,\ H_0(\cdot ))\\ 
&H_0\left(\psi_1,\psi_2\right)=Gamma(\psi_1,\psi _2). 
\end{split}
\end{equation*}
The prior for the seasonal dependency parameter $\thebeta \sim \mathcal{N}({\mu }_0,{\sigma }^2_0)$. The reason for not placing a DP over $\thebeta$ is that, we believe the claim numbers are driven more by peril type than seasonal patterns. Finally, the summarized Dirichlet Prior-Hierarchical Bayesian Collective Risk Model (DP-HBCRM) can be written as
\begin{equation*}
\begin{split}
& S_{i,t} \vee n_{i,t} , \kappa_i, \theta_i \sim Inv. Gamma\left(n_{i,s}\cdot \kappa_i, \theta_i\right), \quad  \theta_i>0, \kappa_i>0,\\ 
& N_{i,t} \vee \lambda_{i,s}  \sim Poisson\left(\lambda_{i,s}\right),\quad \lambda_{i,s}>0,\\ 
& \log \left(\lambda_{i,s} \right) \vee \alpha_i, \thebeta, X_s = \alpha_i+ \thebeta X_s,\\
& (\kappa_i, \theta_i) \sim \ DP(\gamma_1,\ G_0(\cdot )),\\
& \alpha_i \sim \ DP(\gamma_2,\ H_0(\cdot )),\\ 
& \thebeta \sim \mathcal{N} (\mu _0,\sigma^2_0),\\ 
& G_0\left(\zeta_1,\zeta_2,\eta_1,\eta_2\right)=Gamma(\zeta_1, \zeta _2)\times Gamma(\eta_1,\eta_2),\\
& H_0\left(\psi_1, \psi_2\right)=Gamma(\psi_1,\psi_2), 
\end{split}
\end{equation*}
with the following distributions for its hyperparameters
\begin{equation*}
\begin{split}
&{\zeta }_1\sim \ Gamma(0.01,0.01), \enskip {\zeta }_2\sim \ Gamma(0.01,0.01), \enskip{\eta }_1\ \sim \ Gamma(0.01,0.01),\\
&{\eta }_2\ \sim \ Gamma(0.01,0.01),\enskip {\psi }_1\ \sim \ Gamma(0.01,0.01),\enskip {\psi }_2\ \sim \ Gamma(0.01,0.01)\\
&\thebeta\ \sim \ \mathcal{N}(0,0.01). 
\end{split}
\end{equation*}
The explicit solution for the likelihood function and posterior density for the above expressions is mathematically intractable in closed form due to the introduction of the DP prior. Therefore, we utilize the Markov Chain Monte Carlo (MCMC) techniques to simulate the posterior distribution of model parameter as well as predictive inference. We use Open BUGS (\cite{A48}) to implement the model. 

\subsection{Parameter Calibration of the DP-HBCRM} %2.3 

\subsubsection{Data} %2.3.1\\ 

\begin{figure}[!h]
\centering
\begin{tabular}{cc}
\includegraphics*[width=0.45\textwidth, height=0.25\textheight]{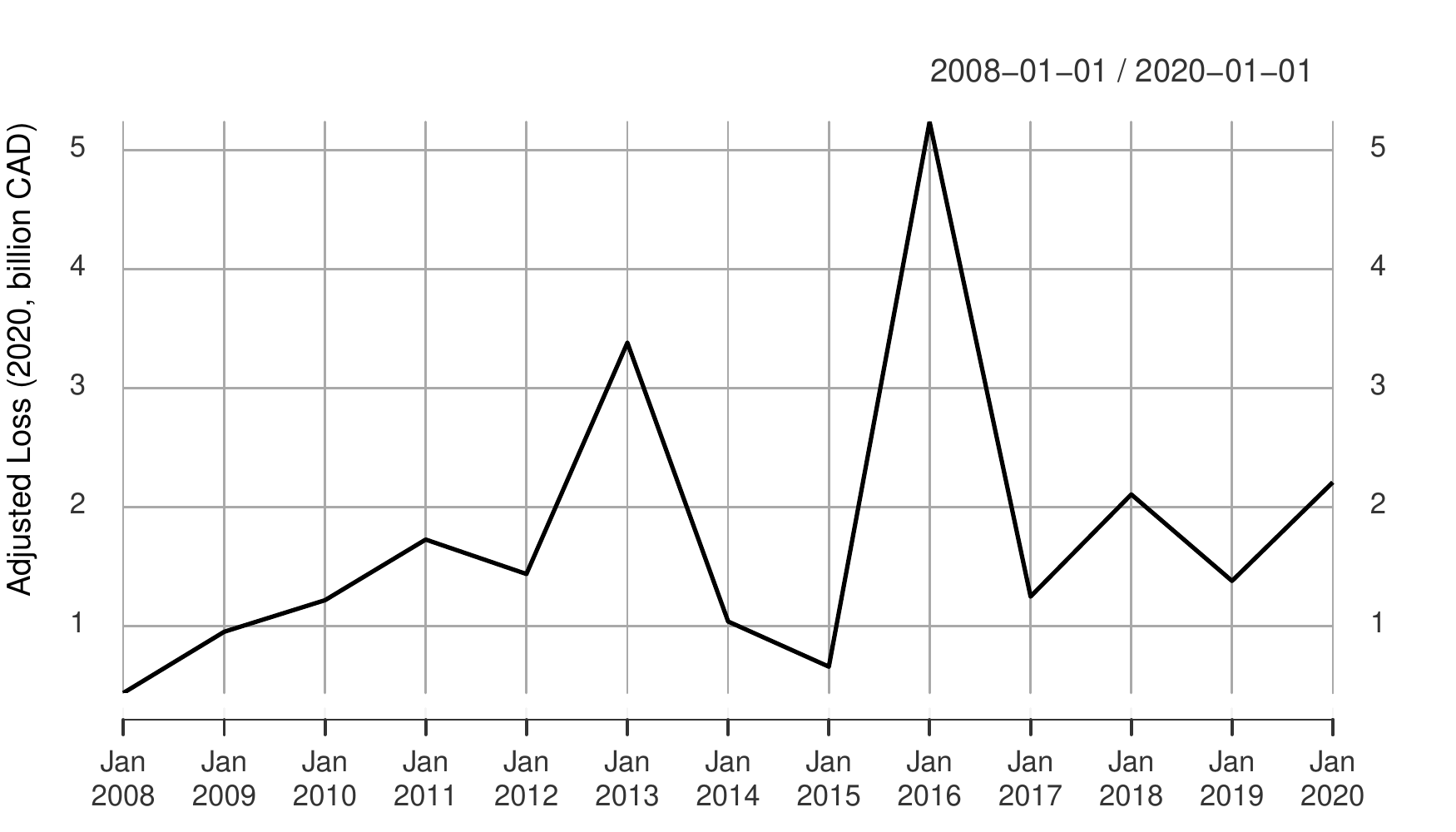} & \includegraphics*[width=0.45\textwidth, height=0.25\textheight]{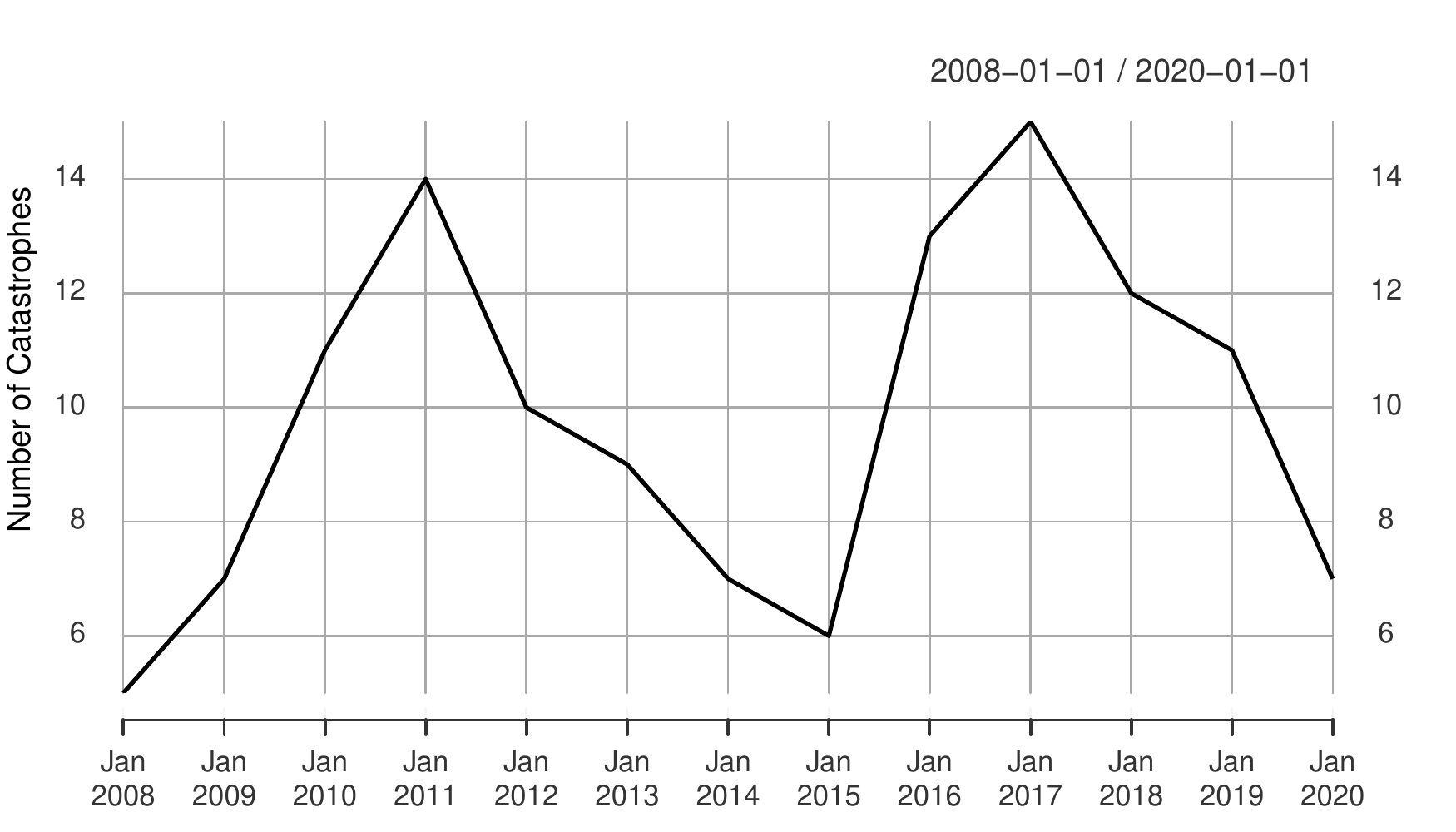}
\end{tabular}
\caption{CatIQ annual catastrophe cost incurred (left) and the number of catastrophes (right) in Canada from 2008 -- 2020.}
\label{Figure2.2}
\end{figure}

In practice, an insurance company keeps records on claims reported by policyholders, such as the number of claims, peril type, claim amount, and peril event dates. Therefore, the aggregate claim $S$, the total number of claims $N$ are observed, hence treated as the model's inputs. This empirical study is conducted using data provided by the Canadian company -- Catastrophe Indices and Quantification Inc. (CatIQ), a subsidiary of PERILS AG in Switzerland.\footnote{\href{https://public.catiq.com}{https://public.catiq.com}} CatIQ is an insurance service company that collates catastrophe loss data from (re)insurers and builds comprehensive industry insured loss and exposure indices to serve the needs of the (re)insurance industries and other stakeholders. CatIQ only includes catastrophe losses to the industry of at least CA\$25 million and affecting a significant number of insurers and policyholders. CatIQ is an authority on insured property losses for Canada. As such, most underwriters use the CatIQ loss index as a veritable source. Our data set on insured catastrophe consists of 130 catastrophes in Canada within the period 2008 -- 2020. Figure \ref{Figure2.2} shows the adjusted total annual CatIQ loss and number of qualified catastrophes between 2008 and 2020. 
%Figure 2.2: 

%2.3.2 
\subsubsection{Bayesian Inference via MCMC}

We run 40,000 steps of MCMC sampling, discard the first 10,000 steps of the chain as burn-in, and use the remaining 30,000 steps for our estimation. Table \ref{tab:posterior} below reports the posterior means of parameters.

\begin{table}[!h]
\centering
\begin{tabular}{|p{0.9in}|p{0.9in}|p{0.8in}|p{0.5in}|p{0.4in}|p{0.5in}|p{0.5in}|p{0.4in}|} \hline 
\vspace{-0.5cm}\center\textbf{Peril} & \multicolumn{7}{|p{3.6in}|}{\vspace{-0.5cm}\center\textbf{Posterior means}} \\ \hline 
 & \multicolumn{2}{|p{1.4in}|}{\vspace{-0.5cm}\center ${S}_{{i}{\ }}$\textbf{}} & \multicolumn{5}{|p{2.2in}|}{\vspace{-0.5cm}\center ${{N}}_{{i},{s}{\ }}$\textbf{}} \\ \hline 
 & ${{\kappa }}_{{i}}$\textbf{} & ${{\theta }}_{{i}}$\textbf{} & ${{\lambda }}_{{i},{1}}$\textbf{} & ${{\lambda }}_{{i},{2}}$\textbf{} & ${{\lambda }}_{{i},{3}}$\textbf{} & ${{\lambda }}_{{i},{4}}$\textbf{} & ${{\alpha }}_{{i}}$\textbf{} \\ \hline 
Windstorm\textbf{} & ${1.268}\times{{10}}^{{-}{4}}$ & ${8.076\times}{{10}}^{{-}{9}}$ & ${10,168}$ & ${10,813}$ & ${11,499}$ & ${12,229}$ & ${9.165}$ \\ \hline 
Severe Storm\textbf{} & ${1.226}\times{{10}}^{{-}{4}}$ & ${7.586\times}{{10}}^{{-}{9}}$ & ${19,400}$ & ${20,632}$ & ${21,941}$ & ${23,334}$ & ${9.811}$ \\ \hline 
Hailstorm \textbf{} & ${1.268}\times{{10}}^{{-}{4}}$ & ${8.086\times}{{10}}^{{-}{9}}$ & ${12,160}$ & ${12,932}$ & ${13,753}$ & ${14,626}$ & ${9.344}$ \\ \hline 
Winter Storm \textbf{} & ${1.268}\times{{10}}^{{-}{4}}$ & ${8.086\times}{{10}}^{{-}{9}}$ & ${18,243}$ & ${19,401}$ & ${20,633}$ & ${21,942}$ & ${9.750}$ \\ \hline 
Flood \textbf{} & ${1.859}\times{{10}}^{{-}{4}}$ & ${1.420\times}{{10}}^{{-}{8}}$ & ${16,119}$ & ${17,142}$ & ${18,231}$ & ${19,388}$ & ${9.626}$ \\ \hline 
Tornado\textbf{} & ${1.617}\times{{10}}^{{-}{4}}$ & ${1.142\times}{{10}}^{{-}{8}}$ & ${11,901}$ & ${12,656}$ & ${13,460}$ & ${14,314}$ & ${9.323}$ \\ \hline 
Hurricane\textbf{} & ${1.583}\times{{10}}^{{-}{4}}$ & ${1.139\times}{{10}}^{{-}{8}}$ & ${18,243}$ & ${19,401}$ & ${20,633}$ & ${21,942}$ & ${9.750}$ \\ \hline 
Tropical Storm  & ${2.948}\times{{10}}^{{-}{3}}$ & ${3.259\times}{{10}}^{{-}{7}}$ & ${14,777}$ & ${15,715}$ & ${16,712}$ & ${17,773}$ & ${9.539}$ \\ \hline 
Fire  & ${4.095}\times{{10}}^{{-}{3}}$ & ${4.572\times}{{10}}^{{-}{7}}$ & ${9,077}$ & ${9,653}$ & ${10,266}$ & ${10,918}$ & ${9.052}$ \\ \hline 
\end{tabular}

\caption{Posterior means of the DP-HBCRM.}
\label{tab:posterior}
\end{table}\newpage 

The induced clusters in the aggregate claims parameters are rather self-explanatory. For instance, perils 1, 2, 3, and 4 have somehow similar parameter estimates. Perils 5, 6, 7 also show some similarities in terms of their parameter estimates, while perils 8 and 9 exhibit very different parameter estimates. Peril types show a strong differentiation in the number of claims. We note that the number of claims is greatly influenced by ${\alpha }_i$, the random intercept in nonhomogeneous Poisson regression. The seasonality component $\thebeta=0.0615$ is not zero; hence it contributes significantly to the claim number process. Next, we report the group indicator and group frequency for each peril in Table \ref{tab:distribution}. %MFD => already mentioned: The Bayesian outputs are obtained based on 30,000 MCMC values after a burn-in of 10,000.

%Table 2.3:
 
\begin{table}[!h]
\centering %p{0.9in}|p{3.4in}|p{1.9in}
\includegraphics*[width=\textwidth]{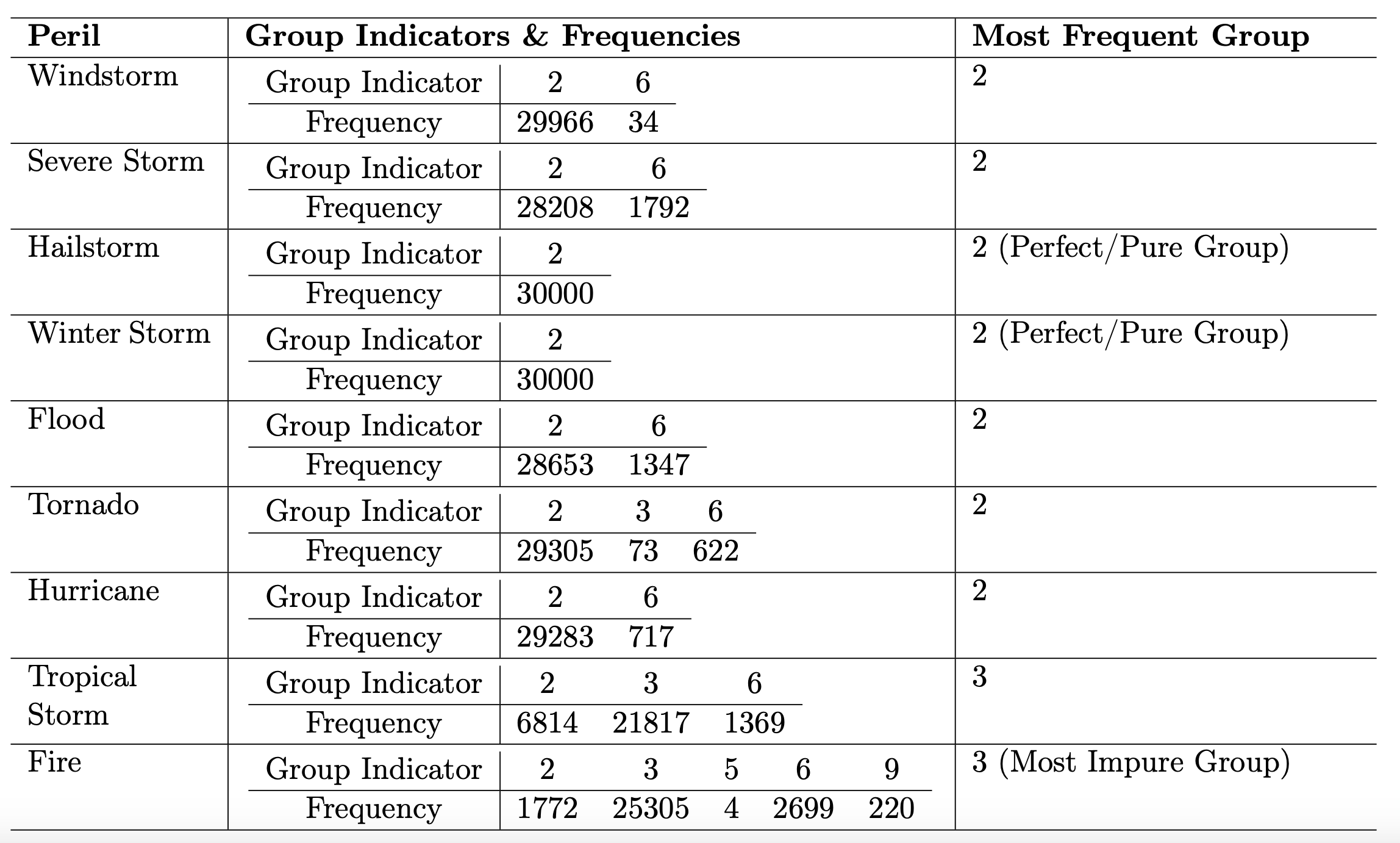}
\caption{Distribution of $30,000$ MCMC samples under various aggregate claim $(S_i)$ groups induced by the DP-HBCRM.}
\label{tab:distribution}
\end{table}\newpage 

Hailstorm and Winter Storm show pure grouping by assigning the same group indicator to all MCMC samples. For example, in the case of Hailstorm, we observe 100\% occupancy for Cluster 2. This is also true for Winter Storm. Therefore, according to the DP-HBCRM, Hailstorm and Winter Storm can be grouped. Windstorm, Severe Storm, Flood, Tornado, and Hurricane also report more than 98\% occupancy for Cluster 2, leading them to be grouped in Cluster 2. Tropical Storm and Fire seem to show higher occupancy in Cluster 3 (about 72.7\% for peril 8 and 84.3\% for peril 9). We separate these two perils due to the  marked difference observed in their claim numbers (see Table \ref{tab:posterior}). Our data shows that fire occurs with a relatively low frequency but has the highest claim cost per occurrence. The DP-HBCRM captured this behavior. Based on the above analysis, we group the aggregate claim parameters in four clusters, as reported in Table \ref{tab:distinct}. The clustering based on the number of claims, $N_i$ is presented in Table~\ref{1A} in Appendix~\ref{appendix}.
%Table 2.4: 
\begin{table}[!h]
\centering
\begin{tabular}{|p{1.5in}|p{3.7in}|} \hline 
\textbf{Claims clusters} & \textbf{Perils} \\ \hline 
Cluster 1 & Windstorm, Severe storm, Flood, Tornadoes, Hurricane \\ \hline 
Cluster 2 & Hailstorm, Winter storm, \\ \hline 
Cluster 3 & Tropical storm \\ \hline 
Cluster 4 & Fire \\ \hline 
\end{tabular}
\caption{Distinct clusters of aggregate claim parameters.}
\label{tab:distinct}
\end{table}

%2.3.3 
\subsubsection{Convergence Diagnostics}

Markov chains need to converge before we can reliably interpret results from them. Due to the stochasticity of the DP-HBCRM model, we cannot directly assess the convergence of the aggregate claims per se. Instead, we test for convergence on the hyperparameters, $\left({\zeta }_1,{\zeta }_2,{\eta }_1,{\eta }_2\right)$  of the base distribution $G_0$. The assumption that, when parameters under the base measure converges, then the target distribution converges can be supported by the \textit{weak }approximation property of the DP probability measure. \\

We perform two checks based on graphical techniques. We ran three parallel Markov chains with 50,000 steps each, including a 10,000 burn-in period. Each of the chains simulated in parallel has different starting points, permitting verification on whether the sequences mix, indicating convergence. As reported in Figure \ref{fig:trace} of Appendix A, the three chains mix well, indicating convergence. The kernel density plot shows no multimodal distributions, further corroborating convergence. A more robust way to verify convergence is to perform the Brooks-Gelman-Rubin (BGR) test (\cite{A19}). The BGR test compares the variance with and between the chains. We report the BGR test in Figure \ref{fig:BGR} of Appendix A. 

%Figure 2.3:

\subsubsection{Bayesian Predictive Inference via MCMC.} %2.3.4 

The goal of the catastrophic risk model developed is to predict the probability that a catastrophe would occur with aggregate claim cost being more than a stipulated threshold. To put it more formally, consider the number of the claims process, $\left\{N_t\ :\ t\in [0,T]\right\}$ using the nonhomogeneous Poisson process (NHPP) with parameters $\lambda \left(t\right)>0$ where, for avoidance of doubt, $T$ is now the maturity and not the number of quarters. Then the probability of aggregate claims, $S_t$ being less than or equal to a threshold, $D$ at time $t$ is the marginal distribution of aggregate claim obtained by convolution as 
\begin{equation}\label{eq:2.3}
S_t(D)=\sum^{\infty }_{n=0}{e^{-\lambda \left(T-t\right)}\frac{{\left(\lambda \left(T-t\right)\right)}^n}{n!}}F^{n*}\left(D\right).
\end{equation}
We should note that Equation~\eqref{eq:2.3} is the explicit form of  Equation~\eqref{eq:2.2}. In classical approaches, to approximate the density function of $S\left(t\right)$ involves convolution methods that often require complicated and lengthy numerical integration. An alternative approach under the Bayesian method is to use MCMC simulation.\\

We learn about the distribution of $S(t)$, by using all the posterior samples of its parameters, $\phi =(\lambda ,\kappa ,\theta )$. Consider the posterior predictive distribution of some variable $z$ (which is dependent on $\phi $) given observed data $Y$ defined by 
%\MFD{$\pi$ is used for the prior and posterior, and $p$ is the likelihood, here and must be elsewhere too.}
\begin{equation*}
\pdf \left(z\ \mathrel{\left|\vphantom{z\  \ Y}\right.\kern-\nulldelimiterspace}\ Y\right)=\int_{\phi }{p\left(z\ \right|\ \phi )\pdf (\phi \ |}\ Y)d\phi 
\end{equation*}
where $\pdf \left(\phi \ \right|\ Y)$ is the posterior distribution of $\phi $ given the data $Y$, and $p\left(z\ \right|\ \phi )$ is the density of $z$ given $\phi $. By integrating over all possible values of $\phi $, we account for variability in parameter estimates, often ignored in classical approaches. Consequently, we obtain better estimates than if we had used point estimates such as the posterior mean. Since we cannot integrate directly over $\phi $, we instead use the MCMC method to obtain posterior predictive distribution estimates. Let $\left(t,\ T\right),\ t\ge 0$ be a time interval of interest for which we want to predict aggregate claims. Given $N$ number of past claims, we can predict the future number of claims, $N_f$ as
\begin{eqnarray*}
 \prob\left[N_f=n \vee N\right]&=&\int^{\infty }_0\prob\left[N_f=n\ \left|\ \theta ,\ N\right]\pdf \left(\theta \ \right|\ N\right)d\theta\\
 & =&\mathbb{E}_{\theta \vee N}\bigg[\prob\left[N_f=n \vee \theta \right]\bigg]\\
& = & \mathbb{E}_{\theta \vee N}\left[\frac{e^{-\theta }{\theta }^n}{n!}\right]= \sum^{\infty }_{n=0}{\frac{e^{-{\lambda }^{(i)}(T-t)}{\left({\lambda }^{(i)}\left(T-t\right)\right)}^n}{n!}},\enskip \text{as} \enskip N_f\ \sim \ Poisson\left({\lambda }^{(i)}_s\right),\\
& \approx&  \frac{1}{m}\sum^m_{i=1}{\frac{e^{-{\lambda }^{(i)}(T-t)}{\left({\lambda }^{(i)}\left(T-t\right)\right)}^n}{n!}},
\end{eqnarray*}
where $m$ is the number of MCMC simulated samples and ${\lambda }^{(i)}$ is the \textit{i}${}^{th}$ simulated value of $\left(\theta \ \right|\ N)$. Similarly, given $S$ past aggregate claims, we predict future claims, $S_f$ as follows

\begin{equation*}
\begin{split}
&\prob\left[S\ \le S_f \vee n_f,\ u\right]=\int_u\prob\left[S\ \le\ S_f\ |\ U=u\ ,\ N_f=n_f\right]du\\
& {=\mathbb{E}}_{u\vee s}\bigg[\prob\left[S\le S_f \vee\ \kappa,\ \theta,\ n_f\right]\bigg]\approx \frac{1}{m}\sum^m_{i=1}{{\left\{1-{\left(\frac{{\theta }^{(i)}}{S_f}\right)}^{{n_f\cdot \kappa }^{(i)}}\right\}}^{-1}},
\end{split}
\end{equation*}
where $u\ \sim  Inv.Gamma(\kappa ,\theta )$, ${\kappa }^{(i)}$ and ${\theta }^{(i)}$ are the \textit{i}${}^{th}$ simulated values $\kappa $ and $\theta $ respectively. We simulate  $\ \prob\left[S\le D\right]$  at time interval $\left(t,\ T\right),\ t\ge 0$ in the following steps: 

\begin{enumerate}
\item  Obtain estimates ${\lambda }^{(i)},\ {\kappa }^{(i)},\ {\theta }^{(i)}$ for the parameters $\lambda ,\ \ \kappa ,\ \ \theta.$

\item  Simulate $N_f\ \sim \ Poisson\left({\lambda }^{\left(i\right)}\right).$

\item  Simulate $N_f$ values of $S\ \sim \ Inv.\ Gamma(N_f\cdot {\kappa }^{\left(i\right)},\ {\theta }^{\left(i\right)}).$

\item  Evaluate the probability that $S^{\left(i\right)}\le D$.
\end{enumerate}

\section{Stochastic Interest Rate Model}\label{sec3} %3.0

\subsection{CIR Model} %3.1

\noindent 

The instantaneous interest (short rate) dynamics proposed by \cite{A5} assume a mean-reverting process. In this model, the diffusion term has a `square root'. The model is a benchmark for obtaining analytical bond prices. Our choice for using the CIR model over others such as the Vasicek model is that our interest rate data have positive interest rates. The latter admits negative interest rates, which is not suitable in this paper. The short rate dynamics $\left\{r_t~:~t\ \in [0,T]\right\}$ is defined by the following stochastic differential equation
\begin{equation}\label{eq:3.1} 
dr_t=\left(\alpha -\beta r_t\right)dt+~\sigma \sqrt{r_t}dW_t,
\end{equation}
with the condition  $\alpha >{\sigma }^2/2$ to ensure that $r_t\geq 0$.  $\left\{W_t:~t\ge 0\right\}$ is a standard Brownian motion and $\alpha ,~\beta ,~\sigma ~\ge 0$ are parameters to be estimated. The estimation of the continuous-time model in Equation \eqref{eq:3.1} from discretely observed data can be problematic since the likelihood is difficult to obtain (see, e.g., \cite{A27}). It is  a well known fact that the conditional density of the CIR process follows a non-central chi-squared distribution. However, it may not always be possible to get this distribution in practice. In this paper, we discretize the continuous-time model using the Euler-Maruyama approximation scheme as 
\begin{equation}
r_{t+\Delta t}=r_t+\left(\alpha -~\beta r_t\right)\Delta t +~\sigma \sqrt{\Delta t}\sqrt{r_t}{\epsilon }_t,~~~\Delta t=\frac{T}{N},~~\Delta t>0,
\end{equation}
where $\Delta t$ is the time interval and $\epsilon_{\mathrm{t}}~\mathrm{\sim }~\mathcal{N}\left(\mathrm{0,1}\right).$ 

\subsubsection{Data Augmentation} %3.1.1 

Some difficulty arises in the use of discretization when the magnitude of the observation interval $\Delta t=t_{i+1}-t_i$ is large. We overcome this difficulty in the Bayesian approach by introducing augmented data between each pair of observations. Suppose, with some abuse of notation, we have $T$ observations, and $M$ augmented data between each pair of consecutive observations available:
\begin{equation*}
t_i={\tau }_{0,i}<{\tau }_{1,i}<\dots <{\tau }_{M,i}=t_{i+1}.
\end{equation*}
The stepsize of the augmented data ${\delta }_{\tau }={\tau }_{j+1,i}-{\tau }_{j,i}$ is kept small enough to ensure the accuracy and stability of the Euler-Maruyama scheme. Let $Y=(r_1,r_2,\dots ,\ r_T)$ denote all observations and $Y^*=(r^*_1,r^*_2,\dots ,\ r^*_{T-1})$ be augmented data where ${r}^*_t=(r^*_{t,0},r^*_{t,1},\dots ,r^*_{t,M})$ and $r^*_{t,0}=r_t$. For each time $t\ge 0$, we define $\Delta =\frac{\Delta t }{M+1}$ and assume that $r^*_{t,j}$ is a Markov process for $j=0,1, \dots, M.\ $\\
We assume that the transition density of the CIR model follows a Markov process:
\begin{equation*}
p\left(r^*_t\mathrel{\left|\vphantom{r^*_t Y,\alpha ,~\beta ,~\sigma }\right.\kern-\nulldelimiterspace}Y,\alpha ,~\beta ,~\sigma \right)=p\left(r^*_{t,0},r^*_{t,1},\dots ,r^*_{t,M}\mathrel{\left|\vphantom{r^*_{t,0},r^*_{t,1},\dots ,r^*_{t,M} Y,\alpha ,~\beta ,~\sigma }\right.\kern-\nulldelimiterspace}Y,\alpha ,~\beta ,~\sigma \right)=\prod^M_{j=1}{p(}r^*_{t,1}\left|r^*_{t,j-1},\ \alpha ,~\beta ,~\sigma \right),
\end{equation*}
where $r^*_{t,0}=r_t$. In this case the full conditional transition density of the process is 
\begin{equation*}
r^*_{t,j+1}|r^*_{t,j},\alpha ,~\beta ,~\sigma \ \sim \ N(r^*_{t,j}+\left(\alpha -\beta r^*_{t,j}\right)\mathrm{\Delta },\ {\sigma }^2\mathrm{\Delta }r^*_{t,j}).
\end{equation*}

%\subsection{}

\subsection{Bayesian CIR Model} %3.2

\subsubsection{Bayesian CIR Model Specification} %3.2.1 

\noindent 

To arrive at parameter estimates, we need to find the likelihood and full conditional posterior distribution of the CIR model in closed form. We derive the fully conditional posterior distribution by dividing the parameters $(\alpha ,~\beta ,~\sigma )$ into $(\Psi,\ {\sigma }^2)$ where $\Psi={\mathrm{(}\mathrm{\alpha }\mathrm{,}\mathrm{\beta }\mathrm{)}}^T$. This is possible since $(\alpha ,~\beta )$ is linear in the drift component so that we can estimate them jointly as $\Psi ={\mathrm{(}\mathrm{\alpha }\mathrm{,}\mathrm{\beta }\mathrm{)}}^T$(see, e.g., \cite{A13}). We specify the likelihood and priors of the model as follows:
\begin{equation*}
\begin{split}
& Y^*,\ Y|{\mathrm{\Psi },\ \sigma \ }^2\sim \ MVN\left(\mu ,\ {\mathrm{\Lambda }}^{-1}\right)\\
&{\mathrm{\Psi }}\ \sim {\ Truncated\ MVN}_{(0,\infty )}({\mu }_0,\ {\mathrm{\Sigma }}^{-1}_0) \\
&{\sigma }^2\ \sim \ IG({\upsilon }_0,{\beta }_0),
\end{split}
\end{equation*}
where the likelihood follows a multi-variate normal (MVN) distribution due to the two parameters ${{\mathrm{\Psi}},\ \sigma \ }^2$. 

The prior of ${\mathrm{\Psi}}$\textbf{ }is drawn from a truncated MVN to ensure $\alpha ,~\beta $ remain positive. Finally, the prior of ${\sigma }^2$ is drawed from an inverse gamma (IG) distribution. Due to the independent structure between $(\boldsymbol{\mathrm{\Psi }},\ {\sigma }^2)$, the likelihood function can be written as
\begin{equation*}
\begin{split}
 p(Y,\ Y^*|{\mathrm{\ }}{\mathrm{\Psi }},\ {\sigma }^2)&=\prod^{T-1}_{t=1}{\prod^M_{j=0}{p(}}r^*_{t,j+1}|r^*_{t,j},\ {\mathrm{\Psi }},\ {\sigma }^2)\\
 &\propto exp\left\{\sum^{T-1}_{t=1}{\sum^M_{j=0}{\frac{{-\left\{r^*_{t,j+1}-\left[r^*_{t,j}+\left(\alpha -\beta r^*_{t,j}\right)\Delta \right]\right\}}^2}{2{\sigma }^2\Delta r^*_{t,j}}}}\right\} \\
& \propto \ exp\left\{-\sum^{T-1}_{t=1}{\sum^M_{j=0}{\frac{\left\{{\left[\left(\alpha -\beta r^*_{t,j}\right)\Delta +r^*_{t,j}\right]}^2-2r^*_{t,j+1}\left(\alpha -\beta r^*_{t,j}\right)\Delta \right\}}{2{\sigma }^2\Delta r^*_{t,j}}}}\right\}.
\end{split}
\end{equation*}
After expanding and simplifying, we get
\begin{equation}\label{eq:3.2}
\propto exp\left\{-\frac{\Delta {A\alpha }^2+\Delta B{\beta }^2-2\Delta \left(T-1\right)\left(M+1\right)\alpha \beta -2C\alpha -2D\beta }{2{\sigma }^2}\right\}
\end{equation}
where
\begin{equation*}
\begin{split}
& A=\sum^{T-1}_{t=1}{\sum^M_{j=0}{\frac{1}{r^*_{t,j}}}}, \quad  B=\sum^{T-1}_{t=1}{\sum^M_{j=0}{r^*_{t,j}}}, \\
& C=-\sum^{T-1}_{t=1}{\sum^M_{j=0}{\frac{r^*_{t,j}-r^*_{t,j+1}}{r^*_{t,j}}}}, \quad 
D=\sum^{T-1}_{t=0}{\sum^M_{j=0}{\left(r^*_{t,j}-r^*_{t,j+1}\right)}}.
\end{split}
\end{equation*}
Since the likelihood, $p(Y,\ Y^*|{\mathrm{\ }}{\mathrm{\Psi }},\ {\sigma }^2)$ is a bivariate normal distribution of the terms ${\mathrm{\Psi }},\ {\sigma }^2$, it can be written as
\begin{equation}\label{eq:3.3}
\begin{split}
& MVN\left(\mu ,\ {\mathrm{\Lambda }}^{-1}\right)\propto {\left|{{\mathrm{\Lambda }}}_{{\mathrm{\Psi }}}\right|}^{\frac{1}{2}}exp\left\{-\frac{1}{2}{\left(\boldsymbol{\mathrm{\Psi }}\boldsymbol{\mathrm{-}}{\boldsymbol{\mu }}_{\boldsymbol{\mathrm{\Psi }}}\right)}^T{\mathrm{\Lambda }}^{-1}_{\mathrm{\Psi }}\left(\boldsymbol{\mathrm{\Psi }}\boldsymbol{\mathrm{-}}{{}{\mu }}_{{}{\mathrm{\Psi }}}\right)\right\},
\end{split}
\end{equation}
where ${{}{\mu }}_{{}{\mathrm{\Psi }}}=\left(\genfrac{}{}{0pt}{}{{\mu }_1}{{\mu }_2}\right)$ and ${{}{\mathrm{\Lambda }}}_{{}{\mathrm{\Psi }}}=\left( \begin{array}{cc}
a_{11} & a_{12} \\ 
a_{21} & a_{22} \end{array}
\right)$. Then it follows that
\begin{equation}\label{eq:3.4}
\begin{split}
& \quad {\left({}{\mathrm{\Psi }}{}{\mathrm{-}}{{}{\mu }}_{{}{\mathrm{\Psi }}}\right)}^T{\mathrm{\Lambda }}_{\mathrm{\Psi }}\left({}{\mathrm{\Psi }}{}{\mathrm{-}}{{}{\mu }}_{{}{\mathrm{\Psi }}}\right) \\
& =\left(\alpha -{\mu }_1,\beta -{\mu }_2\right)
\left( \begin{array}{cc}
a_{11} & a_{12} \\ 
a_{21} & a_{22} \end{array}
\right)\left(\genfrac{}{}{0pt}{}{\alpha -{\mu }_1}{\beta -{\mu }_2}\right)\\
& = a_{11}{\alpha }^2+a_{22}{\beta }^2+2a_{12}\alpha \beta -2\left(a_{11}{\mu }_1+a_{12}{\mu }_2\right)\alpha \\
& -2\left(a_{22}{\mu }_2+a_{12}{\mu }_1\right)\beta +\left(2a_{12}{\mu }_1{\mu }_2+a_{11}{\mu }^2_1+a_{22}{\mu }^2_2\right).
\end{split}
\end{equation}
Comparing formula~\eqref{eq:3.3} and~\eqref{eq:3.4} , we get
\begin{equation*}
\begin{split}
& a_{11}=\frac{\Delta }{{\sigma }^2}A; \quad a_{22}=\frac{\Delta }{{\sigma }^2}B; \quad a_{12}=\frac{\Delta }{{\sigma }^2}\left(T-1\right)\left(M+1\right),\\ & 
a_{11}{\mu }_1+a_{12}{\mu }_2=C; \quad  a_{22}{\mu }_2+a_{12}{\mu }_1=D.
\end{split}
\end{equation*}
Therefore,
\begin{equation*}
{{}{\mu }}_{{}{\mathrm{\Psi }}}={\left(\frac{a_{22}C-a_{12}D}{a_{11}a_{22}-a^2_{12}},\frac{-a_{12}C+a_{11}D}{a_{11}a_{22}-a^2_{12}}\right)}^T
\end{equation*}
and
\begin{equation*}
{{}{\mathrm{\Lambda }}}_{{}{\mathrm{\Psi }}}=\left( \begin{array}{cc}
\frac{\Delta }{{\sigma }^2}A & -\frac{\Delta }{{\sigma }^2}\left(T-1\right)\left(M+1\right) \\ 
-\frac{\Delta }{{\sigma }^2}\left(T-1\right)\left(M+1\right) & \frac{\Delta }{{\sigma }^2}B \end{array}
\right).
\end{equation*}

\noindent The full conditional posterior distribution for ${}{\mathrm{\Psi }}$ then becomes 
\begin{equation}\label{eq:3.5}
\pdf\left({}{\mathrm{\Psi }}\mathrel{\left|\vphantom{{}{\mathrm{\Psi }} Y,Y^*,{\sigma }^2}\right.\kern-\nulldelimiterspace}Y,Y^*,{\sigma }^2\right)\propto p(Y,\ Y^*|{}{\mathrm{\ }}{}{\mathrm{\Psi }},\ {\sigma }^2)\ \pdf ({}{\mathrm{\Psi }})
\end{equation}
where
\begin{equation}\label{eq:3.6}
\begin{split}
& \hspace{-1cm}\pdf \left({}{\mathrm{\Psi }}\right)=\ {\left|{\mathrm{\Sigma }}_0\right|}^{\frac{1}{2}}\ \mathrm{exp}\left\{-\frac{1}{2}{\left({\mathrm{\Psi }}_0\mathrm{-}{\mu }_0\right)}^T{\mathrm{\Sigma }}^{-1}_0\left({\mathrm{\Psi }}_0\mathrm{-}{\mu }_0\right)\right\}\\
&\propto \mathrm{exp}\left\{-\frac{1}{2}{\left({\mathrm{\Psi }}_0-{\mu }_0\right)}^T{{\mathrm{\Sigma }}_0}^{-1}\left({\mathrm{\Psi }}_0-{\mu }_0\right)\right\}\\
& =-\frac{1}{2}\mathrm{exp}\left\{{\left({\mathrm{\Psi }}_0-{\mu }_0\right)}^T\left({\mathrm{\Psi }}_0-{\mu }_0\right){{\mathrm{\Sigma }}_0}^{-1}\right\}\\
& \propto -\frac{1}{2}\mathrm{exp}\left\{{{\mathrm{\Sigma }}_0}^{-1}{\mathrm{\Psi }}^T_o{\mathrm{\Psi }}_0-{{\mathrm{\Sigma }}_0}^{-1}{\mathrm{\Psi }}^T_o{\mu }_0-{{\mathrm{\Sigma }}_0}^{-1}{\mu }^T_0{\mathrm{\Psi }}_0+{{\mathrm{\Sigma }}_0}^{-1}{\mu }^T_0{\mu }_0\right\} \\
&  \propto -\frac{1}{2}\mathrm{exp}\left\{{\mathrm{\Psi }}^T_o{{\mathrm{\Sigma }}_0}^{-1}{\mathrm{\Psi }}_0-2{\mathrm{\Psi }}^T_o{{\mathrm{\Sigma }}_0}^{-1}{\mu }_0+{\mu }^T_0{{\mathrm{\Sigma }}_0}^{-1}{\mu }_0\right\} =-\frac{1}{2}\left\{{\mathrm{\Psi }}^T_o{{\mathrm{\Sigma }}_0}^{-1}{\mathrm{\Psi }}_0-2{\mathrm{\Psi }}^T_o{{\mathrm{\Sigma }}_0}^{-1}{\mu }_0\right\}\\
&\propto -\frac{1}{2}\mathrm{exp}\left\{{\mathrm{\Psi }}^T_oA_0{\mathrm{\Psi }}_0-2{\mathrm{\Psi }}^T_ob_0\right\},
\end{split}
\end{equation}
and where $A_0={{\mathrm{\Sigma }}_0}^{-1}$ and $b_0={{\mathrm{\Sigma }}_0}^{-1}{\mu }_0$.\\\\
The multi-variate normal likelihood of equation~\eqref{eq:3.3} can be rewritten as
\begin{equation}\label{eq:3.7}
\begin{split}
& p(Y,\ Y^*|{}{\mathrm{\ }}{}{\mathrm{\Psi }},\ {\sigma }^2)\ \propto {\left|\mathrm{\Lambda }\right|}^{\frac{1}{2}}\ \mathrm{exp}\left\{-\frac{1}{2}{\left(r_i\mathrm{-}\mu \right)}^T{\mathrm{\Lambda }}^{-1}\left(r_i\mathrm{-}\mu \right)\right\}\\
& \propto -\frac{1}{2}\mathrm{exp}\left\{\sum_i{{\mathrm{r}}^T_i{\mathrm{\ }\mathrm{\Lambda }}^{-1}{\mathrm{\ r}}_i}-2\sum_i{{\mu }^T{\mathrm{\ }\mathrm{\Lambda }}^{-1}}{\mathrm{r}}_i+\sum_i{{\mu }^T}{\mathrm{\Lambda }}^{-1}\mu \right\}\propto -\frac{1}{2}\mathrm{exp}\left\{-2{\mu }^T{\mathrm{\ }\mathrm{\Lambda }}^{-1}n\overline{r}+n{\mu }^T{\mathrm{\ }\mathrm{\Lambda }}^{-1}\mu \right\}\\
&=-\frac{1}{2}\mathrm{exp}\left\{-2{\mu }^Tb_1+{\mu }^TA_1\mu \right\},
\end{split}
\end{equation}
where $b_1=n\overline{r}{\mathrm{\ }\mathrm{\Lambda }}^{-1}$ and  $A_1=n{\mathrm{\ }\mathrm{\Lambda }}^{-1}$.\\
Taking the product of Equation~\eqref{eq:3.6} and Equation~\eqref{eq:3.7} gives
\begin{equation*}
\begin{split}
& p(Y,\ Y^*|{}{\mathrm{\ }}{}{\mathrm{\Psi }},\ {\sigma }^2)\ \pdf ({}{\mathrm{\Psi }})\propto -\frac{1}{2}\mathrm{exp}\left\{-2{\mu }^Tb_1+{\mu }^TA_1\mu \right\}\times -\frac{1}{2}\mathrm{exp}\left\{{\mathrm{\Psi }}^T_oA_0{\mathrm{\Psi }}_0-2{\mathrm{\Psi }}^T_ob_0\right\}\\
& \propto \mathrm{exp}\left\{{\mu }^Tb_1-\frac{1}{2}{\mu }^TA_1\mu -\frac{1}{2}{\mu }^TA_0\mu +{\mu }^Tb_0\right\}\propto \mathrm{exp}\left\{{\mu }^T\left(b_1+b_0\right)-\frac{1}{2}{\mu }^T\left(A_0+A_1\right)\right\},
\end{split}
\end{equation*}

\noindent with $A_n=A_0+A_1={{\mathrm{\Sigma }}_0}^{-1}+n{\mathrm{\ }\mathrm{\Lambda }}^{-1}$; $b_n=b_0+b_1={{\mathrm{\Sigma }}_0}^{-1}{\mu }_0+n\overline{r}{\mathrm{\ }\mathrm{\Lambda }}^{-1}$, the full conditional posterior can be simplified as 

\begin{equation}
\pdf\left({}{\mathrm{\Psi }}\mathrel{\left|\vphantom{{}{\mathrm{\Psi }} Y,Y^*,{\sigma }^2}\right.\kern-\nulldelimiterspace}Y,Y^*,{\sigma }^2\right)\propto \mathrm{exp}\left\{{\mu }^Tb_n-\frac{1}{2}{\mu }^T\mu A_n\right\}\propto MVN\left({}{\mathrm{\Psi }}\mathrel{\left|\vphantom{{}{\mathrm{\Psi }} {{\mathrm{\Sigma }}_0}^{-1}{\mu }_0+n\overline{r}{\mathrm{\ }\mathrm{\Lambda }}^{-1},\ {{\mathrm{\ }\mathrm{\Sigma }}_0}^{-1}+n{\mathrm{\ }\mathrm{\Lambda }}^{-1}}\right.\kern-\nulldelimiterspace}{{\mathrm{\Sigma }}_0}^{-1}{\mu }_0+n\overline{r}{\mathrm{\ }\mathrm{\Lambda }}^{-1},\ {{\mathrm{\ }\mathrm{\Sigma }}_0}^{-1}+n{\mathrm{\ }\mathrm{\Lambda }}^{-1}\right).
\end{equation}
The full conditional posterior distribution for ${\sigma }^2$\textbf{ }on the other hand can be expressed as
\begin{equation}
\pdf\mathrm{(}{\sigma }^2\mathrm{|}Y,\ Y^*,{}{\mathrm{\Psi }}\mathrm{)}\propto p(Y,\ Y^*|{}{\mathrm{\ }}{}{\mathrm{\Psi }},\ {\sigma }^2)\ \pdf ({\sigma }^2).
\end{equation}
Therefore, 
\begin{equation*}
\begin{split}
& \pdf\left({\sigma }^2\mathrel{\left|\vphantom{{\sigma }^2 Y,\ Y^*,{}{\mathrm{\Psi }}}\right.\kern-\nulldelimiterspace}Y,\ Y^*,{}{\mathrm{\Psi }}\right)\propto \ exp\left\{\sum^{T-1}_{t=1}{\sum^M_{j=0}{\frac{{-\left\{r^*_{t,j+1}-\left[r^*_{t,j}+\left(\alpha -\beta r^*_{t,j}\right)\Delta \right]\right\}}^2}{2{\sigma }^2\Delta r^*_{t,j}}}}{\sigma }^2\right\}{\left({\sigma }^2\right)}^{-({\upsilon }_0+1)}exp\left[\frac{-{\beta }_0}{{\sigma }^2}\right] \\
& \propto {\left({\sigma }^2\right)}^{-\left({\upsilon }_0+\frac{\left(T-1\right)\left(M+1\right)}{2}+1\right)}\ exp\left\{-\frac{1}{2}\left[\sum^{T-1}_{t=1}{\sum^M_{j=0}{\left(\frac{{-\left\{r^*_{t,j+1}-\left[r^*_{t,j}+\left(\alpha -\beta r^*_{t,j}\right)\Delta \right]\right\}}^2}{r^*_{t,j}}\right)}}+{\beta }_0\right]\right\} \\
& \propto Inverse\ Gamma\left({\sigma }^2\mathrel{\left|\vphantom{{\sigma }^2 {\upsilon }_0+\frac{\left(T-1\right)\left(M+1\right)}{2},\ {\beta }_0+\sum^{T-1}_{t=1}{\sum^M_{j=0}{\frac{{\left\{r^*_{t,j+1}-\left[r^*_{t,j}+\left(\alpha -\beta r^*_{t,j}\right)\Delta \right]\right\}}^2}{2r^*_{t,j}}}}}\right.\kern-\nulldelimiterspace}{\upsilon }_0+\frac{\left(T-1\right)\left(M+1\right)}{2},\ {\beta }_0+\sum^{T-1}_{t=1}{\sum^M_{j=0}{\frac{{\left\{r^*_{t,j+1}-\left[r^*_{t,j}+\left(\alpha -\beta r^*_{t,j}\right)\Delta \right]\right\}}^2}{2r^*_{t,j}}}}\right).
\end{split}
\end{equation*}

\subsubsection{Sampling Algorithm} %3.2.2 
We obtain samples of the model parameters using a Gibbs sampler. The Gibbs sampler is an MCMC algorithm that generates realizations from an assumed distribution for each parameter in turns, conditional on the current values of the other parameters and observed data points. The sequence of samples generated constitutes a Markov chain whose stationary distribution is the joint posterior distribution we desire. It is relatively simpler to sample from the fully conditional posterior distribution (\textit{if available in closed form}) than to marginalize over a joint distribution by integration. Our goal is to simulate samples for $\alpha ,\ \ \beta $ and ${\sigma }^2$ via a Gibbs sampler from the fully conditional posterior distribution derived above, using the following steps:

\noindent \textbf{Step 1}: Initialize $\ {}{\mathrm{\Psi }},{\sigma }^2,r^*_{1,0}$

\noindent \textbf{Step 2}: Use data augmentation to generate samplings of ${{}{Y}}^{{}{*}}{}{=}r^*_1,r^*_2,\dots ,\ r^*_{T-1}$

\noindent \textbf{Step 3}: Use Gibbs sampler to

\begin{enumerate}
\item  update ${}{\mathrm{\Psi }}$ from $\pdf({}{\mathrm{\Psi }}{}{\mathrm{|}}{}{Y},{{}{Y}}^{{}{*}},\ {\sigma }^2)$ where ${{}{Y}}^{{}{*}}{}{=}r^*_1,r^*_2,\dots ,\ r^*_{T-1}$ are from the previous iteration

\item  update ${\sigma }^2$ from $\pdf({\sigma }^2|{}{Y},{{}{Y}}^{{}{*}},{}{\mathrm{\Psi }})$ where ${{}{Y}}^{{}{*}}{}{=}r^*_1,r^*_2,\dots ,\ r^*_{T-1}\ $are from the prior iteration, and ${}{\mathrm{\Psi }}$ is from (a)
\end{enumerate}

\noindent \textbf{Step 4}: Update $r^*_1,r^*_2,\dots ,\ r^*_{T-1}$ from $p(r^*_t|{}{\mathrm{\Psi }},{\sigma }^2{}{\mathrm{)\ }}$\textbf{}

\noindent \textbf{Step 5}: Repeat Step 3 and Step 4 until the sampling size \textit{N} is reached.

\paragraph{Forecasting} % 3.2.3 

\noindent Given the MCMC samples of each parameter $\{{\alpha }^{\left(i\right)},{\beta }^{\left(i\right)}\ and\ {\sigma }^{\left(i\right)},\ i=1,\dots ,N\}$ and an initial interest, $r_0$, we can forecast future interest rate $\{r^{\left(i\right)}_1,r^{\left(i\right)}_2,\dots ,r^{\left(i\right)}_T,\ i=1,\dots ,N\}$ recursively as

\begin{equation*}
\begin{split}
& r^{(i)}_1=r_0+\left({\alpha }^{(i)}-~{\beta }^{(i)}r_0\right)\Delta +~{\sigma }^{(i)}\sqrt{\Delta }\sqrt{r_0}{\epsilon }^{(i)}_1 \\
& r^{\left(i\right)}_2=r^{(i)}_1+\left({\alpha }^{\left(i\right)}-~{\beta }^{\left(i\right)}r^{(i)}_1\right)\Delta +~{\sigma }^{\left(i\right)}\sqrt{\Delta }\sqrt{r^{(i)}_1}{\epsilon }^{\left(i\right)}_2 \\
& \vdots \\ 
& r^{\left(i\right)}_T=r^{(i)}_{T-1}+\left({\alpha }^{\left(i\right)}-~{\beta }^{\left(i\right)}r^{(i)}_{T-1}\right)\Delta +~{\sigma }^{\left(i\right)}\sqrt{\Delta }\sqrt{r^{(i)}_{T-1}}{\epsilon }^{\left(i\right)}_T,
\end{split}
\end{equation*}
where ${\epsilon }^{\left(i\right)}_t\sim \ N\left(0,\ 1\right),\ i=1,\dots ,N,\ \ t=1,\dots ,T$.

\subsection{Parameter Calibration of the Bayesian CIR Model} % 3.3 
\begin{figure}[!h]
\centering
\includegraphics*[width=0.8\textwidth, height=0.3\textheight]{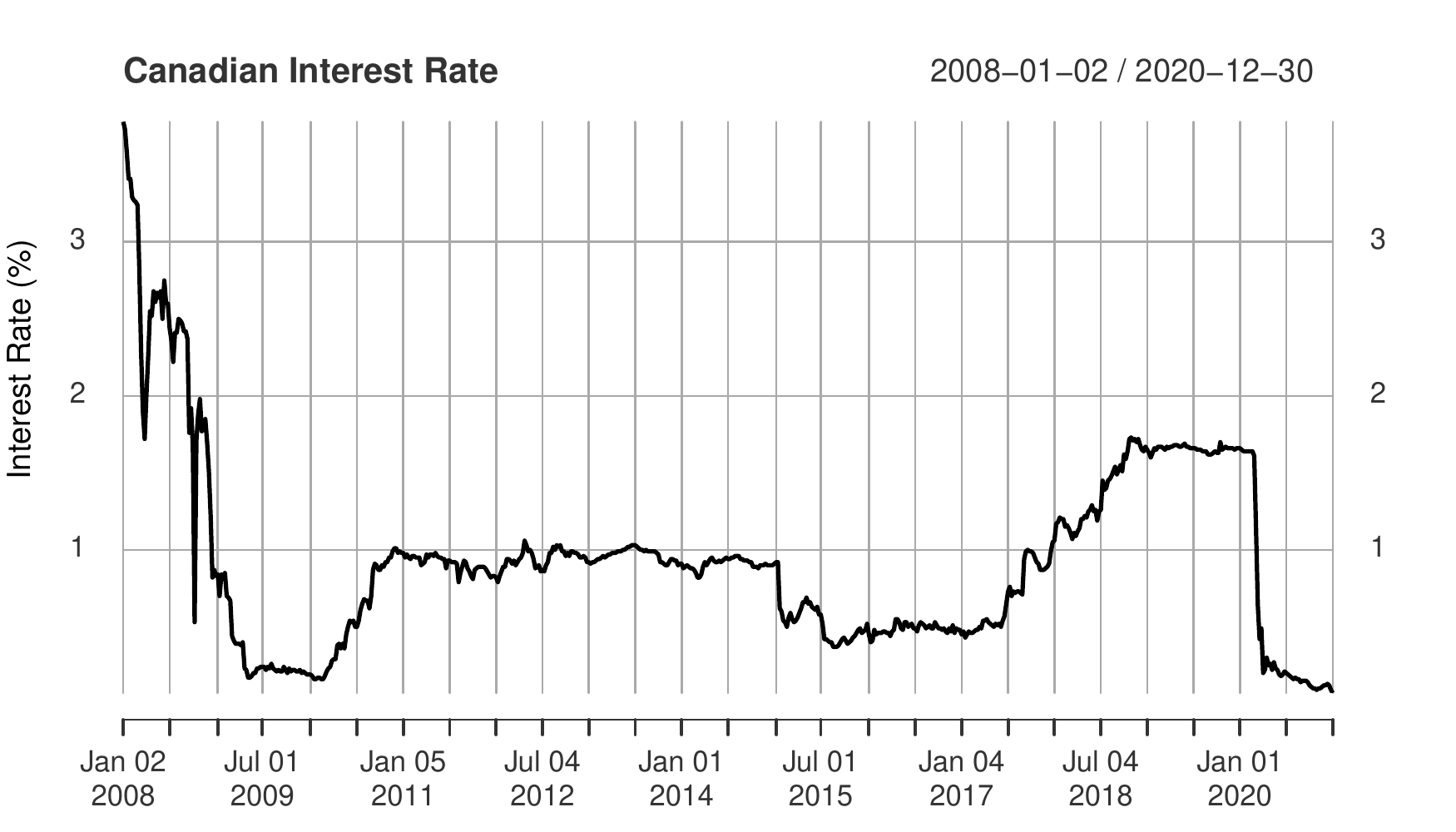}
\caption{Time series of Canadian 3-month Treasury bills from 2008 to 2020.}
\label{fig:t-bills}
\end{figure}

We calibrate the parameters of the CIR model using 12 years of weekly data on 3-months Canadian Treasury bills from January 2, 2008, to December 12, 2020\footnote{ Due to the independence between interest and catastrophe risk it is not necessary to use a data set covering the same period in time.}. Figure \ref{fig:t-bills} shows the historical yields on the Canadian 3-month Treasury bills. 

%Figure 3.1: 

\noindent We run 15,000 steps of MCMC sampling with $M=20$ and $\Delta =1/252$. We discard the first 5,000 steps of the chain and use the remaining 10,000 steps for our estimation. Table \ref{tab:hyper} below reports the specifications of hyperparameters used in our model.

%Table 3.1: 
\begin{table}[!h]
    \centering
 {\renewcommand{\arraystretch}{1.2}%
  \begin{tabular}{|p{1.2in}|p{0.9in}|} \hline 
Hyperparameter & Specification \\ \hline 
${\upsilon }_0$ & 2.1 \\ \hline 
${\beta }_0$ & 3.0 \\ \hline 
${\mu }^1_0$ & 0.0 \\ \hline 
${\mu }^2_0$ & 0.0 \\ \hline 
${{\mathrm{\Sigma }}_0}^{-1}$ & 10.0 \\ \hline 
\end{tabular}}
    \caption{Hyperparameters of the Bayesian CIR model.}
\label{tab:hyper}
\end{table}

\noindent The posterior mean, posterior standard deviation, 95\% highest posterior density (HPD), and Geweke convergence diagnostics are reported in Table \ref{tab:summary}.

%Table 3.2: 
\begin{table}[!h]
    \centering
   \begin{tabular}{|p{0.8in}|p{0.8in}|p{0.8in}|p{1.4in}|p{0.5in}|} \hline 
Parameters & Posterior Mean & SD & 95\% HPD & Geweke \\ \hline 
$\alpha $ & 3.0299 & $2.07\times {10}^{-3}$ & (3.0258, 3.0339) &  0.3852 \\ \hline 
$\beta $ & 3.2694 & $3.56\times {10}^{-3}$ & (3.2625, 3.2765) &  1.3129 \\ \hline 
${\sigma }^2$ & 0.00171 & $1.90\times {10}^{-5}$ & (0.001673, 0.001747) &  0.8433 \\ \hline 
\end{tabular}
    \caption{Summary statistics of MCMC samples.}
    \label{tab:summary}
\end{table}

\noindent The Geweke diagnostic is in favor of the null hypothesis of convergence since all the z-score values are well within two standard deviations of zero. Figure \ref{fig:trace} below displays the trace plots of the MCMC sampling sequence for $\alpha ,\beta $ and ${\sigma }^2$ which may lend additional support to the convergence of each sequence. The long-term interest rate in the CIR model is modeled as $\alpha /\beta $. Our estimated value for the long-term interest rate is $\frac{3.0299}{3.2694}=0.9267\%$ with a deviation of only about 0.0074 from the historical mean of the observed data, $0.9342\%$. The posterior distribution of $\alpha , \beta $ and ${\sigma }^2$ shown in Figure \ref{fig:CIR_params} appear unimodal and skew-free, resembling a Gaussian.

%Figure 3.2: 
\begin{figure}[!h]
    \centering
  \includegraphics*[width=3.94in, height=3.76in]{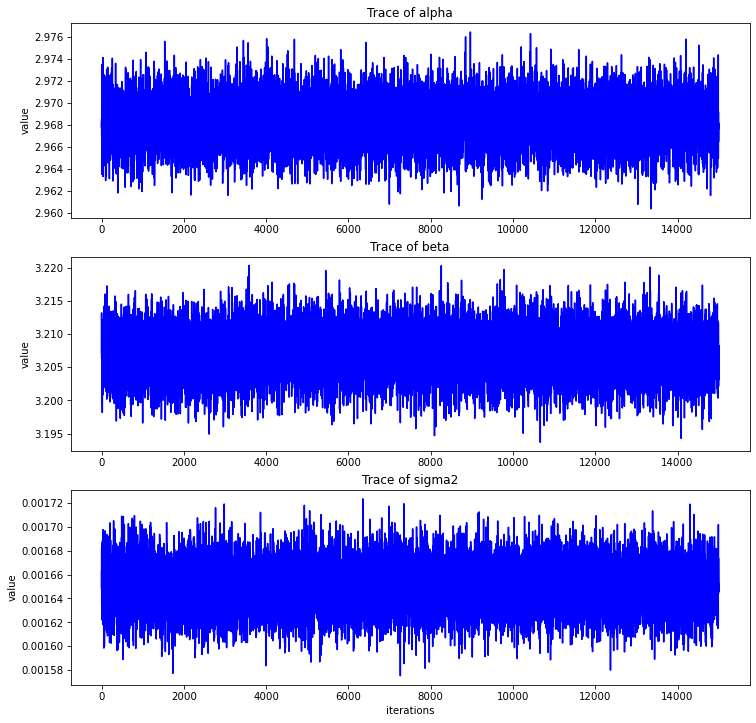}
    \caption{Trace plots of parameters $\alpha, \beta $ and ${\sigma }^2$ from the fully conditional posterior distribution with $M=20$, $N=15,000$.}
    \label{fig:trace}
\end{figure}

%Figure 3.3: 
\begin{figure}[!h]
    \centering
   \includegraphics*[width=5.57in, height=2.90in]{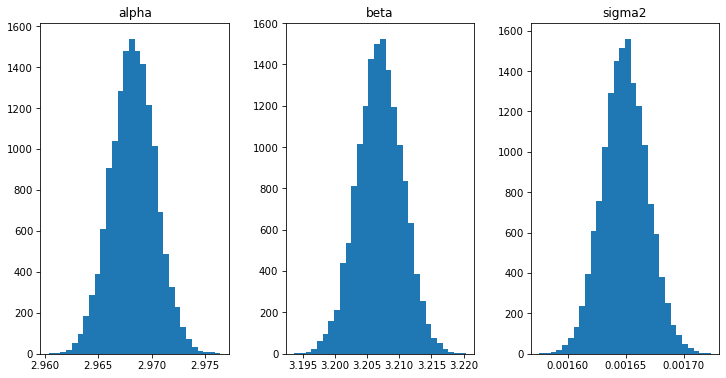}
    \caption{Posterior distribution of Bayesian CIR parameters.}
    \label{fig:CIR_params}
\end{figure}\newpage

\section{Contingent Claim Pricing Model}\label{sec4}

\noindent 

CAT bonds are typically priced under certain assumptions. First, there exists an arbitrage-free market with an equivalent martingale measure. Second, financial markets behave independently of catastrophe occurrence. Third, interest rate changes can be replicated using existing financial instruments. With some abuse of notation, let $0<T<\ \infty $ represent the maturity date for the continuous-time trading interval $\left[0,T\right].$ Then the market uncertainty can be defined by a filtered probability space $(\mathrm{\Omega },\ \mathcal{F},{\left\{{\mathcal{F}}_t\right\}}_{t\in \left[0,T\right]}\mathrm{,}\mathbb{P}\mathrm{)}$, where ${\mathcal{F}}_t$ is a family of $\sigma $-algebras containing all available information about the underlying processes. 

We consider the process that could trigger payment of CAT bond to be a nonnegative non-decreasing, right-continuous stochastic process $Y=\{Y_t,\ t\ \ge 0\}\ $defined on a filtered physical probability space $({\mathrm{\Omega }}^{(1)},\ {\mathcal{F}}^{1},{\left\{{{\mathcal{F}}_t}^{1}\right\}}_{t\in \left[0,T\right]},{\mathbb{P}}^{1})$, where ${{\mathcal{F}}_t}^{1}$ represent all available catastrophic risk information (e.g., claim amount, number of claims, peril type). Consider an arbitrage-free financial market, defined by another filtered physical probability space $({\mathrm{\Omega }}^{(2)},\ {\mathcal{F}}^{2},{\left\{{{\mathcal{F}}_t}^{2}\right\}}_{t\in \left[0,T\right]},{\mathbb{P}}^{2})$, where ${{\mathcal{F}}_t}^{2}$ represents available investment information (e.g., interest rates, inflation, or securities prices). 

Following the argument of \cite{A6} about the independence of the two stochastic processes; we can define a product probability space $(\mathrm{\Omega },\ \mathcal{F},{\left\{{\mathcal{F}}_t\right\}}_{t\in \left[0,T\right]}\mathrm{,}\mathbb{P}\mathrm{)}$, with ${\mathcal{F}}_t=\ {{\mathcal{F}}_t}^{1}\ \times \ {\mathcal{F}}^{2}_t\subset \mathcal{F}$ for $t\ \ge 0$, and ${\mathbb{P}}^{1}\times {\mathbb{P}}^{2}\subset \mathbb{P}$. The independence of the two probability spaces is reasonable considering the low correlation between the underlying catastrophic risks and financial risks. While the latter can be replicated with an available portfolio, the former cannot. This naturally leads us to consider an incomplete market setting for evaluating catastrophe bonds. There is no unique price for a security in an incomplete market setting as there could be infinitely many prices corresponding to different equivalent martingale measures. 

\cite{A39} posits that under the risk-neutral pricing measure $\mathbb{Q}$, the overall economy depends only on financial variables. There is, however, evidence to the contrary that catastrophe risk may pose a substantial systematic risk (see, e.g., \cite{A18}. \cite{A39} argues that the aggregate loss process (i.e,. intensity and severity of losses) retain their original distributional characteristics ${\mathbb{P}}^{1}$ after been transformed from the physical probability measure to the risk-neutral measure ${\mathbb{Q}}^1$ (see for example, \cite{A11}; \cite{A6}; \cite{A31}; \cite{A37}). In other words, the expectation of a random variable $X$ on $\left(\mathrm{\Omega },\ \mathcal{F},{\left\{{\mathcal{F}}_t\right\}}_{t\in \left[0,T\right]}\mathrm{,}\mathbb{P}\right)$ is the same under $\mathbb{Q}$.
\begin{equation*}
\mathbb{E}^{\mathbb{P}}\left[X\right]=\mathbb{E}^{\mathbb{Q}}\left[X\right].
\end{equation*}
To develop a fair pricing model that is consistent with both financial and actuarial valuation, we follow \cite{A51} and apply a distortion function to both physical probability measures ${\mathbb{P}}^{1}$ and ${\mathbb{P}}^{2}$ to transform them to their risk-neutral counterparts ${\mathbb{Q}}^{1}$ and ${\mathbb{Q}}^{2}$. \cite{A51} show that following a distorted probability measure ${\mathbb{Q}}^{1}$, $Y$ becomes heavier tailed than ${\mathbb{P}}^{1}$, hence the riskiness is amplified under ${\mathbb{Q}}^{1}$. One of the most popular risk-neutral approaches is to use the Wang transform proposed by \cite{A53} and \cite{A54}.

\subsection{Pricing Model}

\noindent 

Let us denote the index trigger of a CAT bond as the process, $\{Y_t~:~t\in[0,T]\}$. Then the CAT bond is a derivative product that pays off $\payoff_t \equiv\payoff(Y_t)$ for $t=1,2,\dots,T$ according to a payoff function $\payoff\left(\cdot \right)$, where $\payoff\left(\cdot \right)$ represents a CAT bond whose payments are contingent on $Y_t's\ $process. Let $\{r_t~ :~t\in [0,T]\}$ be an interest rate process. Conditioning on available information on catastrophe risk and financial market performance, the risk-neutral price for a CAT bond which pays $\payoff_t$ at time $t$, $t=1,\dots,T$ is 
\begin{equation}\label{eq:4.1}
\begin{split}
\price_t &=\mathrm{K}\left[\mathbb{E}^{{\mathbb{Q}}^{(2)}}\left[e^{-\int^T_t{r_sds}}\mathrel{\left|\vphantom{e^{-\int^T_t{r_sds}} {\mathcal{F}}^{\left(2\right)}_t}\right.\kern-\nulldelimiterspace}{\mathcal{F}}^{\left(2\right)}_t\right]\cdot\mathbb{E}^{{\mathbb{Q}}^{(1)}}\left[\payoff_T\mathrel{\left|\vphantom{\payoff_T {\mathcal{F}}^{\left(1\right)}_t}\right.\kern-\nulldelimiterspace}{\mathcal{F}}^{\left(1\right)}_t\right]\ \right]\\
&=\mathrm{K}\mathrm{\ }\mathbb{E}^{\mathbb{Q}}\left[\ \left(e^{-\int^T_t{r_sds}}\cdot \payoff_T\mathrel{\left|\vphantom{e^{-\int^T_t{r_sds}}\cdot \payoff_T {\mathcal{F}}_t\ }\right.\kern-\nulldelimiterspace}{\mathcal{F}}_t\ \right)\ \right],
\end{split}
\end{equation}
where $\mathrm{K}$ is the face value, ${\mathcal{F}}_t=\ {{\mathcal{F}}_t}^{1}\ \times \ {\mathcal{F}}^{2}_t\subset \mathcal{F}$ and $\mathbb{Q}=\ {\mathbb{Q}}^{1}\ \times \ {\mathbb{Q}}^{2}$. $\mathbb{E}^{\mathbb{Q}}[\ \cdot |{\mathcal{F}}_t]$ is the expectation of the bond price for the risk-neutral measure. Our task is to statistically model the two independent processes $(Y_t,r_t)$ and convert them to their equivalent risk-neutral measures.\\

There are several trigger approaches to designing payoff structures.  We utilize the industry index trigger to price CAT bonds in this study. According to a CAT bond payment structure, investors receive premiums if the bond is not triggered. This also means that investors may lose their principal if the estimated aggregated loss for the whole industry exceeds a predetermined threshold. CAT bonds can be designed with different payoff functions. We consider one payoff function for a CAT bond contract for illustrative purposes. Assume a zero-coupon CAT bond with maturity date $T$ has a payoff function as follows:
\begin{equation}\label{eq:4.2}
\begin{split}
\payoff=
\begin{cases}
& K, \quad \text{if} \quad  L_T\ \le D, \\
& aK, \quad \text{if}\quad L_T>D,
\end{cases}
\end{split}
\end{equation}
where $L_T$ is a total insured loss at the expiry date $T$, $D$ is the industry index threshold value pre-specified in the bond contract, and $a\in [0,1)$ is a fraction of the principal $K$, which the bondholders must pay when the bond is triggered. Insert payment function of Equation~\eqref{eq:4.2} into Equation~\eqref{eq:4.1} for the price of a zero-coupon bond, and we get

\begin{equation}\label{eq:4.3}
\begin{split}
\price_t &=\mathrm{K}\mathrm{\ }\mathbb{E}^{\mathbb{Q}}\left[\ e^{-\int^T_t{r_sds}}\cdot \payoff_T\mathrel{\left|\vphantom{e^{-\int^T_t{r_sds}}\cdot \payoff_T {\mathcal{F}}_t\ }\right.\kern-\nulldelimiterspace}{\mathcal{F}}_t\ \ \right]\\
& =\mathbb{E}^{\mathbb{Q}}\left[e^{-\int^T_t{r_sds}}\cdot \left(K{\mathbb{I}}_{\left\{L_T\le D\right\}}+a.K{\mathbb{I}}_{\left\{L_T>D\right\}}\right)|{\mathcal{F}}_t\right]\\
& =\mathbb{E}^{\mathbb{Q}}\left[e^{-\int^T_t{r_sds}}\cdot \left(K\cdot Pr[L_T\le D]+a.K\cdot P[L_T>D]\right) |\mathcal{F}_t\right]\\
&{=\mathbb{E}}^{\mathbb{Q}}\left[e^{-\int^T_t{r_sds}}\cdot K\left(F_T(D)+a\left(1-F_T(D)\right)\right)|\mathcal{F}_t\right].  
\end{split}
\end{equation}

\section{Risk-Neutralization of CAT Bond Prices}\label{sec5}

\noindent 

The risk-neutral probability measure $\mathbb{Q}$, is needed to calculate the theoretical CAT price in Equation~\eqref{eq:4.3}. In this paper, to be consistent with our Bayesian pricing framework, we convert the \textit{physical} distributions $(\payoff, r_t)$ into risk-neutral forms using the maximum information entropy proposed by \cite{A50}. Using the maximum entropy approach, we estimate \textit{only one} risk-neutral distribution instead of two separate risk-neutral measures for the processes (i.e., interest and catastrophe risk processes). For a more practical appeal of the entropy approach, see \cite{A33}, \cite{A29}, Li, Kogure, and \cite{A34}. From a Bayesian perspective, the \textit{physical} probability measures can be considered our prior distribution of CAT bond processes. This prior knowledge gets updated to the posterior (risk-neutral) distribution when we observe market prices at time $t=0$.

\subsection{Pricing Mechanism Under the Entropy Principle}

We simultaneously generate $N$ states of the MCMC sampling for the CIR process $\{r_t: t=1,2,\dots ,T\}$ and contingent payoffs $\{\payoff_t: t=1,2,\dots ,T\}$, and denote them by 
\begin{center}
$\left\{\left(r^{(i)}_t,\payoff^{(i)}_t\right):~ i=1,2,\dots ,N,~t=1,2,\dots,T\right\}.\ $
\end{center}
We let $\pi $ denote the \textit{physical} distribution of the \textit{N }states of the MCMC sampling with equal probability of $\frac{1}{N}$ in each state. To implement risk-neutralization, we need at least one market price constraint. Since there are currently no active CAT bonds designed around the CATIQ index, we use a hypothetical risk premium, $\delta =250$ bps above LIBOR, to discount future payoffs to the present as
\begin{equation}\label{eq:5.1}
\price_0=\sum^N_{i=1}{\sum^T_{t=1}{\mathrm{exp}\mathrm{}(-\mathrm{\delta t})}\payoff^{(i)}_t\ {\pi }_i\ },\ \ \ \  
\end{equation}
where $\price_0$ is the issue price of the bond at time $t=0$. Now, the caveat here is that using a hypothetical risk premium for the price constraint does not reflect the accurate market price of risk. However, for this numerical exercise, it may suffice. We convert $\pi $ into the risk-neutral measure ${\pi }^*$ using the market constraint such that
\begin{equation}\label{eq:5.2}
\sum^N_{i=1}{\sum^T_{t=1}{\left(\mathrm{}\mathrm{exp}\left(-\sum^t_{u=1}{r^{(i)}_u}\right)\payoff^{(i)}_t\right){\pi }^*_i}}=\price_0,
\end{equation}
where $\sum^N_{i=1}{\sum^T_{t=1}{\left({\mathrm{exp} \left(-\sum^t_{u=1}{r^{(i)}_u}\right)\ }\payoff^{(i)}_t\right){\pi }^*_i}}=\mathbb{E}^{\mathbb{Q}}\left[\cdot |{\mathcal{F}}_t\right]$\\

This is nothing other than the statement that the expected price at time $t=0$ must equal the payoff at $t=T$, but calculated under the risk-neutral measure instead of the physical probability measure and discounted back using the risk-free rate. Therefore, the risk-neutral approach corrects for risk by adjusting probabilities rather than adjusting the risk-free rate. It corrects physical probabilities to overweight states in which aggregate outcomes are particularly egregious\footnote{There is a close linkage between the risk-neutral measure and the Arrow-Debrau theory under both its no-arbitrage argument and equilibrium conditions (see \cite{A12}).}. Based on the maximum entropy principle, the risk-neutral distribution ${\pi }^*$ should minimize the Kullback-Leibler information divergence
\begin{equation*}
\mathrm{KL}\left(\mathbb{P}\parallel \mathbb{Q}\right)\mathrm{\ :arg\ min\ }\sum^N_{i=1}{{\pi }^*_i\ln\left(\frac{{\pi }^*_i}{{\pi }_i}\right)},
\end{equation*}
with the following additional constraints ${\pi }^*>0,\ for\ i=1,\dots .N,\ \ \ \ \sum^N_{i=1}{{\pi }^*_i\ =1}$. This minimization problem can be solved by the method of Lagrange multipliers as
\begin{equation}\label{eq:5.3}
{\pi }^*_i=\frac{{\pi }_i\mathrm{exp}\left(\lambda \sum^T_{t=1}{\left({\mathrm{exp} \left(-\sum^t_{u=1}{r^{(i)}_u}\right)\ }\cdot \payoff^{\left(i\right)}_t\right)}\right)}{\sum^N_{i=1}{{\pi }_i\mathrm{\ exp}}\left(\lambda \sum^T_{t=1}{\left({\mathrm{exp} \left(-\sum^t_{u=1}{r^{(i)}_u}\right)\ }\cdot \payoff^{\left(i\right)}_t\right)}\right)}\ \ for\ i=1,\dots ,N.
\end{equation}
For brevity of expression, let ${\alpha }_i\triangleq \sum^T_{t=1}{\left({\mathrm{exp} \left(-\sum^t_{u=1}{r^{(i)}_u}\right)\ }\cdot \payoff^{\left(i\right)}_t\right)},$ then Equation~\eqref{eq:5.3} can be simplified as
\begin{equation}\label{eq:5.4}
{\pi }^*_i=\frac{{\pi }_i\mathrm{exp}\left(\lambda {\alpha }_i\right)}{\sum^N_{i=1}{{\pi }_i\mathrm{\ exp}}\left(\lambda {\alpha }_i\right)}\ \ for\ i=1,\dots ,N.
\end{equation}
The Lagrange multiplier, $\lambda$, is given by solving the minimization problem 
\begin{equation}\label{eq:5.5}
\lambda={\mathop{\mathrm{arg\ min\ }}_{\lambda' } \sum^N_{i=1}{\mathrm{exp}\mathrm{}\left[\lambda' \left\{{\alpha }_i-\price_0\right\}\right]}\ }.
\end{equation}
For further details, see Appendix A for the solution and full proof. It follows that the risk-neutral price for a zero-coupon CAT bond can be approximated as
\begin{equation}\label{eq:5.7}
\price_t=\mathrm{K}\mathrm{\ }\mathbb{E}^{\mathbb{Q}}\left[\ e^{-\int^T_t{r_sds}}\cdot \payoff_T\mathrel{\left|\vphantom{e^{-\int^T_t{r_sds}}\cdot \payoff_T {\mathcal{F}}_t\ }\right.\kern-\nulldelimiterspace}{\mathcal{F}}_t\ \ \right]\approx \sum^N_{i=1}{\sum^T_{t=1}{\left({\mathrm{exp} \left(-\sum^t_{u=1}{r^{(i)}_u}\right)\ }\cdot \payoff^{\left(i\right)}_t\right){\pi }^*_i}}
\end{equation}

\subsubsection{Illustrative Example} %5.1.1 

\begin{figure}[!h]
    \centering
    \begin{tabular}{cc}
    \includegraphics*[width=3.21in, height=1.89in]{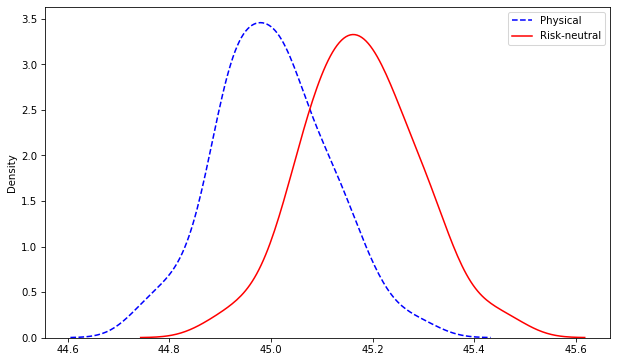} & \includegraphics*[width=3.21in, height=1.88in]{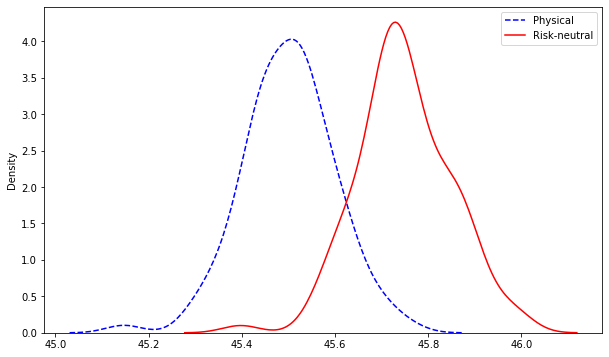}\\
    (a) Multi-Peril Bond Cluster 1  &                           (b) Multi-Peril Bond Cluster 2
    \end{tabular}
    \caption{\textit{Distributions of the present values of multi-peril CAT bonds in clusters 1 and 2.}}
    \label{fig:multi-peril}
\end{figure}

\begin{figure}[!h]
    \centering
    \begin{tabular}{cc}
    \includegraphics*[width=3.24in, height=1.93in]{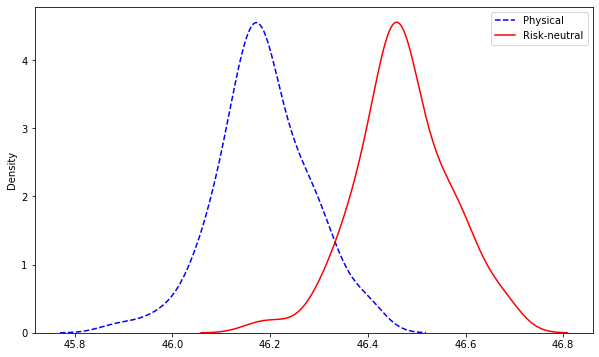} & \includegraphics*[width=3.21in, height=1.89in]{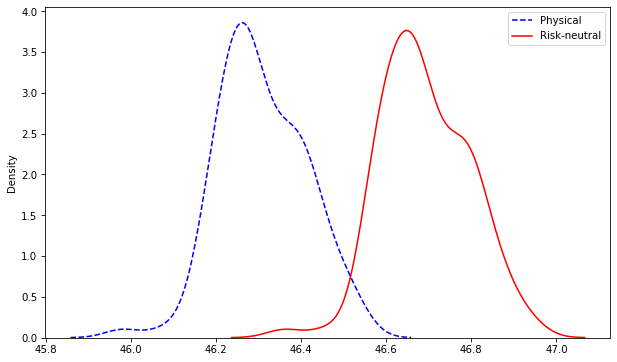}\\
    (a) Single-Peril Bond Cluster 3  &                        (b) Single-Peril Bond Cluster 4
    \end{tabular}
    \caption{\textit{Distributions of the present values of single-peril CAT bonds in clusters 3 and 4.}}
    \label{fig:single-peril}
\end{figure}

We simultaneously draw 10,000 samples from the predictive distributions of DP-HBCRM and CIR models. Using the expression in Equation~\eqref{eq:5.7}, we calculate the present values of a two-year zero-coupon CAT bond with a face value of $K=CA\$100$ belonging to clusters 1 to 4. We use a single threshold $D\in [75.93]$ ten million CAD (3$\times$average annual loss), an initial short-term interest rate, $r_0$ of $0.9267\%$. The results are shown in Figure \ref{fig:multi-peril} for Multi-Peril Cluster 1 and 2 and Figure~\ref{fig:single-peril} for Single-Peril Cluster 3 and 4. The summary statistic of the present values is presented in Table \ref{tab:summary_stats}.

% Figure 5.1: 

% Figure 5.2: 

%Table 5.1: 
\begin{table}[!h]
\centering
\begin{tabular}{|p{0.7in}|p{1in}|p{0.8in}|p{0.7in}|p{0.7in}|p{0.7in}|} \hline 
\textit{} & \textbf{Measure} & \textbf{Mean} & \textbf{SD} & \textbf{Skewness} & \textbf{Kurtosis} \\ \hline 
\textbf{Cluster 1} & Physical & 44.937 & 0.108 & 0.1099 & 0.6331 \\ \hline 
\textbf{} & Risk-neutral & 45.456 & 0.119 & -0.0508 & 0.5991 \\ \hline 
\textbf{Cluster 2} & Physical & 45.50 & 0.097 & 0.2707 & 1.1089 \\ \hline 
\textbf{} & Risk-neutral & 45.75 & 0.100 & -0.0884 & 0.8063 \\ \hline 
\textbf{Cluster 3} & Physical & 46.18 & 0.095 & 0.1359 & 0.6678 \\ \hline 
\textbf{} & Risk-neutral & 46.47 & 0.095 & -0.1342 & 0.5783 \\ \hline 
\textbf{Cluster 4} & Physical & 46.23 & 0.094 & 0.1266 & 0.8413 \\ \hline 
\textbf{} & Risk-neutral & 46.63 & 0.094 & -0.0827 & 0.7338 \\ \hline 
\end{tabular}
\caption{\textit{Summary statistics present values.}}
\label{tab:summary_stats}
\end{table}\newpage 

\paragraph{\textbf{Discussion}}
We note that the expected theoretical price of the CAT bonds across all four clusters is relatively higher in the risk-neutral distribution than the physical one. The results also show that the risk-neutral distribution is more skewed to the left than the physical distribution for all cases,  reflecting the risk adjustment for interest rate and catastrophe risk. We note that there is also an observable positive price difference between Multi-Peril Cluster 1 and Single-Peril Cluster 4 for instance. This implies that indeed Single-Peril Cluster 4 is valued higher than Multi-Peril Cluster 1. While it's difficult to extrapolate this to a general rule, it underscores the relevance of exploring the individual and grouped risk profiles when assessing CAT bond prices designed around an industry index trigger. For a fixed threshold interval and time to maturity, the prices of CAT bonds are affected mainly by the probability of occurrence of a catastrophic event with the duration of the bond contract.\\

Therefore, the aggregate loss, conditional on the number of claims, will be high when there's an expected high number of claims and vice-versa. When aggregate claims are high enough, there is a higher probability of breaching the pre-specified industry threshold, triggering the bond to be paid out to the re-insurer or insurer. Consequently, CAT bonds with higher predicted aggregate claims such as Multi-Peril Cluster 1 have lower expected prices relative to Single-Peril Cluster 4 with lower predicted aggregate claims. In Section \ref{sect:rpa}, we will examine how these price differences affect the risk premia on bonds. 

\subsection{Risk Premium Assessment}\label{sect:rpa}

While the effect of cluster membership in CAT bond prices merits discussion, a more interpretable and actionable representation is to evaluate the risk premia per annum. Thes can easily be derived by solving the root of Equation~\eqref{eq:5.8} for $\delta $. We seek a unique risk premium that matches two discounted payoffs equal so that
\begin{equation}\label{eq:5.8}
\sum^T_{t=1}{\left(\mathbb{E}^{\mathbb{Q}}\left[\ \left(e^{-\int^T_t{r_sds}}\cdot \payoff_T\mathrel{\left|\vphantom{e^{-\int^T_t{r_sds}}\cdot \payoff_T {\mathcal{F}}_t\ }\right.\kern-\nulldelimiterspace}{\mathcal{F}}_t\ \right)\ \right]\ -{\mathrm{exp}\mathrm{}\mathrm{(-}\mathrm{\delta }\mathrm{t)}\mathbb{E}}^{\mathbb{P}}\left[\left(\payoff_T\mathrel{\left|\vphantom{\payoff\_T {\mathcal{F}}_t\ }\right.\kern-\nulldelimiterspace}{\mathcal{F}}_t\ \right)\right]\right)}=0.
\end{equation}

\begin{figure}[!h]
\centering
\includegraphics*[width=4.59in, height=2.98in]{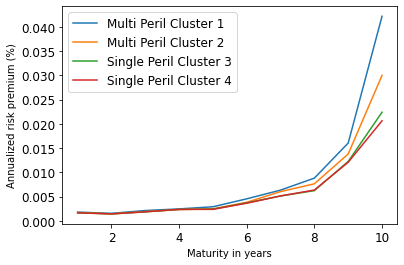}
\caption{Annualized risk premia under stochastic interest rate and DP-HBCRM catastrophe models.}
\label{fig:risk_premia}
\end{figure}

\noindent We calculate the risk premia with maturities ranging from 1 to 10 years, as presented in Figure \ref{fig:risk_premia}. 

%Fig 5.3
Since the risk premium for all maturities is unknown, we assume a hypothetical premium (i.e. $\delta$ = 250 bps) for discounting all future payoffs to give the market price, $\price_0$.
In other words, the market price is the same for all four peril clusters. This aides comparison  of CAT bond prices. However, there may be differences in risk premia for different perils and maturities. From Figure~\ref{fig:risk_premia}, we observe that for a a fixed index trigger threshold, peril-specific CAT bonds exhibit different risk premium profiles. First, all CAT bond clusters show an increase in risk premium as their maturity time increases. This is expected since an investor's risk exposure increases with higher holding periods; hence they must be compensated with a higher risk premium. Risk premia at lower maturities are flat (1 -- 5 years) with no discernible differences across peril clusters but increase sharply afterwards. There exists a non-linear and somewhat exponential relationship between maturity and risk premia. This may explain why most CAT bonds have short maturity dates --- typically between 2 -- 5 years. However, there are some with longer maturities. The short maturity of the bonds somewhat mitigates the risk of losing principal invested. Riskier bonds, i.e., perils with a higher probability of breaching threshold, show higher risk premiums such as Multi-Peril Cluster 1 and vice-versa such as Single-Peril Cluster 4.

\section{Concluding Remarks}\label{sec6}
 The primary contribution of this paper is to present a unified Bayesian CAT bond pricing framework based on uncertainty quantification of catastrophes and interest rates. Our framework allows for complex beliefs about catastrophe risks to capture the distinct and common patterns in catastrophe occurrences, and when combined with stochastic interest rates, yields a unified asset pricing approach with informative expected risk premia. The framework allows us to gain insight into the catastrophe risk profiles of peril-specific bond contracts and, consequently, their expected risk premia and values. Our empirical study contributes to a better understanding of how multi-peril and single-peril CAT bond contracts can be more accurately priced. \footnote{The current ratio of single peril to multi-peril CAT bond outstanding is 4:6 (see, e.g.,\cite{A24}).}\\

Our proposed unified pricing approach remains consistent with financial asset pricing and actuarial valuation theories. The method is shown to be flexible, robust, and incorporates parameter uncertainties in the estimation process. The architecture of the modified Hierarchical Collective risk model, with a Dirichlet prior, is shown to accommodate several characteristics of risk in the calculation of aggregate claims, hence yielding robust estimates. In future work,  construction of the non-homogeneous Poisson process could explore, for example, location-specific factors that can influence aggregate claims. There may be specific geographies that may be more susceptible to hurricanes, for example. 

Another advantage of our approach is the dynamic clustering induced in the posterior samples of parameters. The dynamic clustering ensures that when new information or data is available, these groupings can adjust/change to measure risk correctly. The Bayesian CIR model also offers more precision in the estimation of future interest rates. One major benefit of our maximum entropy approach for calculating the risk premia is that it requires no subjective input from the user, in contrast to other techniques such as the Wang Transform (\cite{A35}). Our entropy approach to risk-neutral pricing can incorporate different market prices, which is especially pertinent in the CAT bond market. \\

Of course, the results provided in this paper have some important limitations which we discuss here.
Chiefly, the flexibility of DP-HBCRM is limited by the number of parameters and prior assumptions on those parameters. An avenue for future research is to consider a three-parameter family of continuous probability distributions. It may enhance the flexibility of our DP-HBCRM for catastrophe risk. Furthermore, a generalized inverse gamma density $f(x;a,d,p)$ may show a different rate of decay for different perils, which can enhance our understanding. 

Other limitations are somewhat ubiquitous in Bayesian financial computing. Our Bayesian CIR is a one-factor model which assumes that the interest rate process is driven by only a single stochastic factor. It is well known that a single factor does not fully capture the term structure dynamics. However, one can easily incorporate several different terms of interest rate into the Bayesian CIR model (see for example \cite{A27}). The computational methods outlined in this study may be computationally expensive, especially in calculating catastrophe probabilities and merits further investigation and potential methodological refinements for parallel computing.

\paragraph{\textbf{Acknowledgments}} 
The authors would like to thank Morton Lane for his valuable comments on this paper.
%This work is from Dickson Nkwantabisa's dissertation under the guidance of Arpita Chatterjee. 

\paragraph{\textbf{Funding}} 
This research did not receive any specific grant from funding agencies in the public, commercial, or not-for-profit sectors.

\bibliographystyle{nonumber}

\bibliography{main}
\newpage
\appendix

\section{Appendix A}\label{appendix}

\noindent 

%\noindent Clustering based on catastrophe counts ($N_i$)

\begin{table}[h!]
\center
\begin{tabular}{|p{1.5in}|p{3.5in}|} \hline 
Peril  & Group indicators and corresponding frequencies. \\ \hline 
1 (Windstorm)  & Group Indicator 3 \newline Frequency 30000 \newline  \\ \hline 
2 (Severe Storm) & Group Indicator \textbf{1} \newline Frequency 30000 \newline  \\ \hline 
3 (Hailstorm)  & \textbf{Group Indicator 9} \newline    Frequency 30000 \newline  \\ \hline 
4 (Winter Storm)  & \textbf{Group Indicator 7} \newline Frequency 30000 \newline  \\ \hline 
5 (Flood)  & \textbf{Group Indicator 6} \newline Frequency 30000 \newline  \\ \hline 
6 (Tornado)  & \textbf{Group Indicator 5} \newline Frequency 30000 \newline  \\ \hline 
7 (Hurricane) & \textbf{Group Indicator 7} \newline Frequency 30000 \newline  \\ \hline 
8 (Tropical Storm)  & \textbf{Group Indicator 8} \newline Frequency 30000 \newline  \\ \hline 
9 (Fire)  & \textbf{Group Indicator          2                    4} \newline Frequency   10560                     19440 \newline  \\ \hline 
\end{tabular}
\caption{Clustering based on catastrophe counts ($N_i$).}
\label{1A}
\end{table}

\begin{figure}[!h]
\centering
\begin{tabular}{cc}
\includegraphics*[width=3in, height=2.1in]{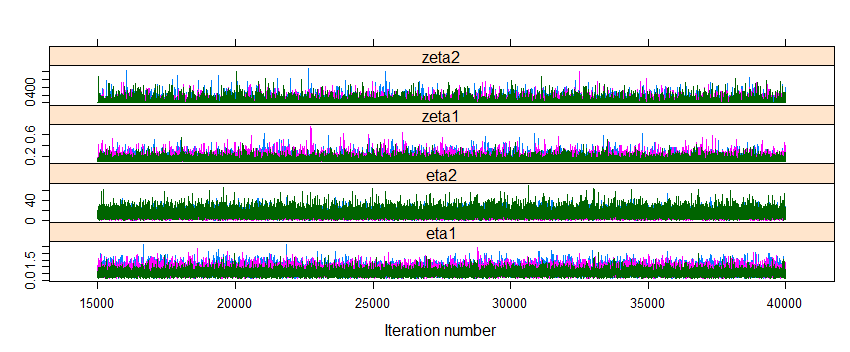} & \includegraphics*[width=3.07in, height=2in]{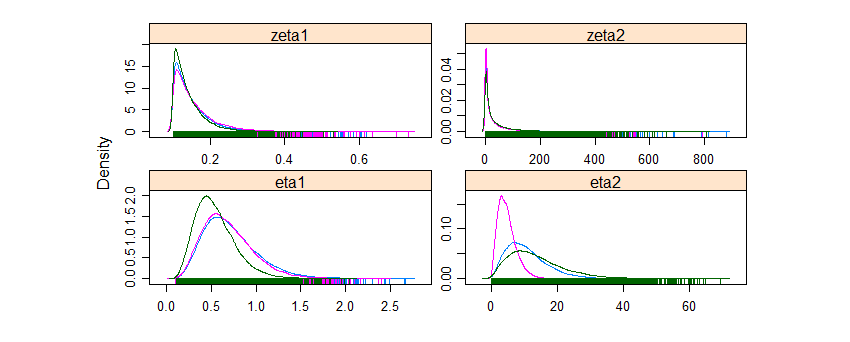}
\end{tabular}
\caption{Trace of MCMC chains and density plots for the base measure on aggregate claims.}
\label{fig:trace}
\end{figure}

%Figure 2.4: 
\begin{figure}[!h]
\centering
    \includegraphics*[width=6in, height=4in]{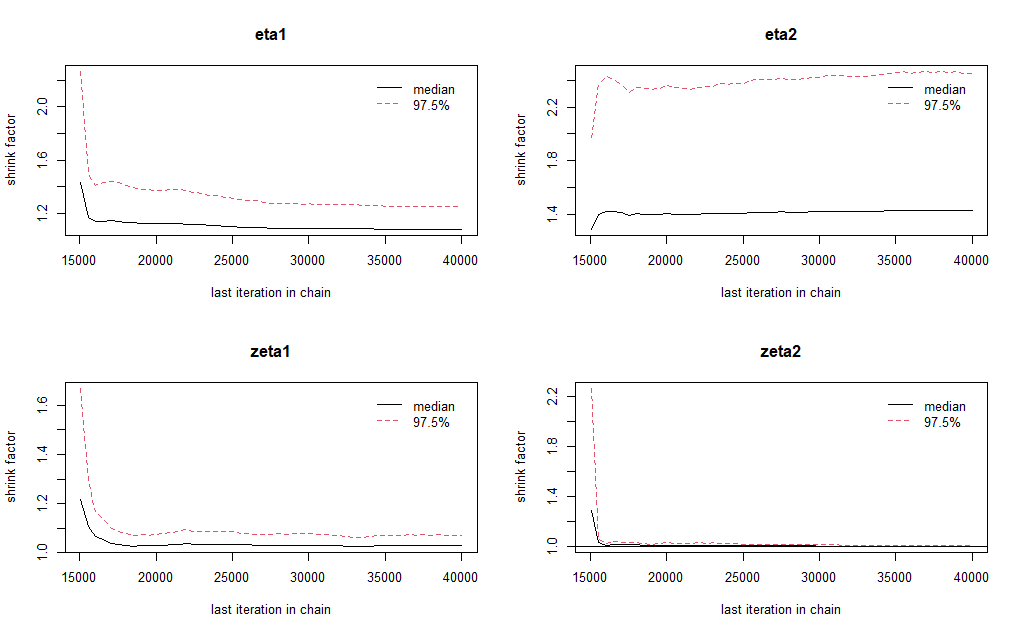}
\caption{BGR convergence plot obtained for the base measure on aggregate claims.}
\label{fig:BGR}
\end{figure}

\clearpage

\subsection{Finding the maximum entropy by the process of Lagrange multipliers}

Based on the maximum entropy principle the risk-neutral distribution ${\pi }^*$ should minimize the Kullback-Leibler information divergence
\begin{equation*}
\mathrm{KL}\left(\mathbb{P}\parallel \mathbb{Q}\right)\mathrm{\ :arg\ min\ }\sum^N_{i=1}{{\pi }^*_i\ln\left(\frac{{\pi }^*_i}{{\pi }_i}\right)}
\end{equation*}
with the following additional constraints ${\pi }^*_i>0,\ for\ i=1,\dots .N,\quad \sum^N_{i=1}{{\pi }^*_i\ =1}$.\\

\noindent This minimization problem can be solved by the method of Lagrange multipliers as follows:
\begin{equation*}
\begin{split}
& \mathcal{L}=\sum^N_{i=1}{{\pi }^*_i\ln\left(\frac{{\pi }^*_i}{{\pi }_i}\right)}-\ \gamma \left(\sum^N_{i=1}{{\pi }^*_i\ -1}\right)-\lambda \left[\left\{\sum^N_{i=1}{\sum^T_{t=1}{\left(\mathrm{}\mathrm{exp}\left(-\sum^t_{u=1}{r^{(i)}_u}\right)\payoff^{(i)}_t\right){\pi }^*_i}}\right\}-\price_0\right]\\
&\gamma,\lambda \in \mathbb{R}.
\end{split}
\end{equation*}
Taking first order conditions with respect to $\gamma $ and $\lambda $, we get
\begin{equation*}
\begin{split}
&\frac{\partial \mathcal{L}}{\partial \gamma }=\ -\gamma \left(\sum^N_{i=1}{{\pi }^*_i\ -1}\right)=0 , \\
&\frac{\partial \mathcal{L}}{\partial \lambda }=\sum^N_{i=1}{\sum^T_{t=1}{\left(\mathrm{exp}\left(-\sum^t_{u=1}{r^{(i)}_u}\right)\payoff^{(i)}_t\right){\pi }^*_i}}=0.
\end{split}
\end{equation*}
\noindent Taking first order conditions with respect to ${\pi }^*_i$, we get

\begin{equation*}
\begin{split}
& \frac{\partial \mathcal{L}}{\partial {\pi }^*_i}=\ln\frac{{\pi }^*_i}{{\pi }_i}+1-\gamma -\lambda \left[\sum^T_{t=1}{\left(\mathrm{exp}\left(-\sum^t_{u=1}{r^{(i)}_u}\right)\payoff^{(i)}_t\right){\pi }^*_i}\right]=0\\
& \mathrm{exp}\left(\ln\frac{{\pi }^*_i}{{\pi }_i}+1-\gamma -\lambda \left[\sum^T_{t=1}{\left(\mathrm{exp}\left(-\sum^t_{u=1}{r^{(i)}_u}\right)\payoff^{(i)}_t\right){\pi }^*_i}\right]\right)=1\\
& \mathrm{exp}\left(\ln{\ \pi }^*_i-{\mathrm{\ln} {\pi }_i\ }+1-\gamma -\lambda \left[\sum^T_{t=1}{\left(\mathrm{exp}\left(-\sum^t_{u=1}{r^{(i)}_u}\right)\payoff^{(i)}_t\right){\pi }^*_i}\right]\right)=1 \\
& {\pi }^*_i={\pi }_i\ \mathrm{exp}\left(-1+\gamma +\lambda \left[\sum^T_{t=1}{\left(\mathrm{exp}\left(-\sum^t_{u=1}{r^{(i)}_u}\right)\payoff^{(i)}_t\right){\pi }^*_i}\right]\right).
\end{split}
\end{equation*}
Expanding to get
\begin{equation}\label{eq:2}
{\pi }^*_i={\mathrm{exp} \left(\mathrm{-}\mathrm{1+}\mathrm{\gamma }\right)\ }{\pi }_i\mathrm{exp}\left(\lambda \sum^T_{t=1}{\left(\mathrm{exp}\left(-\sum^t_{u=1}{r^{(i)}_u}\right)\payoff^{(i)}_t\right)}\right)
\end{equation}
Since $\sum^N_{i=1}{{\pi }^*_i\ \ =1}$ and $\sum^N_{i=1}{{\pi }_i\ \ =1}$, it follows that
\begin{equation*}
\sum^N_{i=1}{{\pi }^*_i=}{\mathrm{exp} \left(\mathrm{-}\mathrm{1+}\mathrm{\gamma }\right)\ }\sum^N_{i=1}{{\pi }_i\mathrm{\ exp}}\left(\lambda \sum^T_{t=1}{\left(\mathrm{exp}\left(-\sum^t_{u=1}{r^{(i)}_u}\right)\payoff^{(i)}_t\right)}\right)=1.
\end{equation*}
Dividing through, we get
\begin{equation}\label{eq:3}
{\mathrm{exp} \left(\mathrm{-}\mathrm{1+}\mathrm{\gamma }\right)\ }\mathrm{=}\frac{1}{\sum^N_{i=1}{{\pi }_i\mathrm{\ exp}}\left(\lambda \sum^T_{t=1}{\left(\mathrm{exp}\left(-\sum^t_{u=1}{r^{(i)}_u}\right)\payoff^{(i)}_t\right)}\right)}
\end{equation}
Inserting equation~\eqref{eq:3} into equation~\eqref{eq:2} gives 
\begin{equation}\label{eq:4}
{\pi }^*_i=\frac{{\pi }_i\mathrm{exp}\left(\lambda \sum^T_{t=1}{\left(\mathrm{exp}\left(-\sum^t_{u=1}{r^{(i)}_u}\right)\payoff^{(i)}_t\right)}\right)}{\sum^N_{i=1}{{\pi }_i\mathrm{\ exp}}\left(\lambda \sum^T_{t=1}{\left(\mathrm{exp}\left(-\sum^t_{u=1}{r^{(i)}_u}\right)\payoff^{(i)}_t\right)}\right)}\ \ \text{for}\ i=1,\dots ,N.
\end{equation}
For brevity of expression, let ${\alpha }_i\triangleq \sum^T_{t=1}{\left(\mathrm{exp}\left(-\sum^t_{u=1}{r^{(i)}_u}\right)\payoff^{(i)}_t\right)}$ . Then equation \eqref{eq:4} can be simplified as
\begin{equation}\label{eq:5}
{\pi }^*_i=\frac{{\pi }_i\mathrm{exp}\left(\lambda {\alpha }_i\right)}{\sum^N_{i=1}{{\pi }_i\mathrm{\ exp}}\left(\lambda {\alpha }_i\right)}\ \ \text{for}\ i=1,\dots ,N.
\end{equation}
To avoid overflow in the numerical calculation of exponents, let $\mathrm{\Gamma }=max\left\{\lambda {\alpha }_i\right\}$. Then equation \eqref{eq:5} is equivalent to
\begin{equation*}
{\pi }^*_i=\frac{{\pi }_i\mathrm{exp}\left(\lambda {\alpha }_i-\mathrm{\Gamma }\right)}{\sum^N_{i=1}{{\pi }_i\mathrm{\ exp}}\left(\lambda {\alpha }_i-\mathrm{\Gamma }\right)}\ \ \text{for}\ i=1,\dots ,N.
\end{equation*}
$\lambda$ is the Lagrange multiplier and can be found in the following steps:\\

\noindent First, we simplify the market constraint in equation \eqref{eq:5.2} as
\begin{equation}\label{eq:6}
\sum^N_{i=1}{{\pi }^*_i}{\alpha }_i=\price_0.
\end{equation}
Substituting equation \eqref{eq:5} into equation \eqref{eq:6}, we get
\begin{equation}
\sum^N_{i=1}{\frac{{\pi }_i\mathrm{exp}\left(\lambda {\alpha }_i\right)}{\sum^N_{i=1}{{\pi }_i\mathrm{exp}\left(\lambda {\alpha }_i\right)}}}{\alpha }_i=\price_0.
\end{equation}
Remembering that ${\pi }_i=\frac{1}{N}$, we can simplify to 
\begin{equation}\label{eq:7}
\frac{\sum^N_{i=1}{{\alpha }_i\mathrm{exp}\left(\lambda {\alpha }_i\right)}}{\sum^N_{i=1}{\mathrm{exp}\left(\lambda {\alpha }_i\right)}}=\price_0.
\end{equation}
This expression for $\price_0$ is the minimizer of the following minimization problem
\begin{equation}\label{eq:8}
{\mathop{\mathrm{arg\ min\ }}_{\lambda } \sum^N_{i=1}{\mathrm{exp}\mathrm{}\left[\lambda \left\{{\alpha }_i-\price_0\right\}\right]}\ }.
\end{equation}
And it can be proved as follows:\\

\noindent Denoting $f\left(\lambda \right)=\sum^N_{i=1}{\mathrm{exp}\mathrm{}\left[\lambda \left\{{\alpha }_i-\price_0\right\}\right]}$, we find the stationary value, then
\begin{equation*}
\begin{split}
&\frac{df}{d\lambda }=\sum^N_{i=1}{\left\{{\alpha }_i-\price_0\right\}\mathrm{exp}\mathrm{}\left[\lambda \left\{{\alpha }_i-\price_0\right\}\right]} \\
&{\mathrm{exp} \left[\mathrm{-}\mathrm{\lambda }\price_0\right]\ }\sum^N_{i=1}{\left\{{\alpha }_i-\price_0\right\}{\mathrm{exp} \left[\lambda {\alpha }_i\right]\ }}=0.
\end{split}
\end{equation*}
That is 
\begin{equation*}
\begin{split}
& \sum^N_{i=1}{\left\{{\alpha }_i-\price_0\right\}{\mathrm{exp} \left[\lambda {\alpha }_i\right]\ }}=0 \\
&\sum^N_{i=1}{{\alpha }_i\mathrm{exp}\left[\lambda {\alpha }_i\right]}=\price_0\sum^N_{i=1}{\mathrm{exp}\left[\lambda {\alpha }_i\right]}.
\end{split}
\end{equation*}
Rearranging gives,
\begin{equation}\label{eq:9}
\frac{\sum^N_{i=1}{{\alpha }_i\mathrm{exp}\left(\lambda {\alpha }_i\right)}}{\sum^N_{i=1}{\mathrm{exp}\left(\lambda {\alpha }_i\right)}}=\price_0.
\end{equation}
Equation~\eqref{eq:9} is the same form as equation~\eqref{eq:7}. So solving the minimization problem in equation \eqref{eq:8} is identical to solving equation \eqref{eq:7}.

\end{document}